\documentclass[useAMS,usenatbib,twocolumn]{aastex61}

\usepackage{mathtools}

\DeclarePairedDelimiter{\abs}{\lvert}{\rvert}

\received{}
\revised{}
\accepted{}
\submitjournal{ApJ}

\shorttitle{Monte Carlo simulations of Photospheric emission}
\shortauthors{Bhattacharya et al.}

\begin{document}


\title{Monte Carlo simulations of Photospheric emission in relativistic outflows}

\correspondingauthor{Mukul Bhattacharya}
\email{mukul.b@utexas.edu (MB)} 

\author{Mukul Bhattacharya}
\affiliation{Department of Astronomy, University of Texas at Austin, Austin, TX 78712, USA}
\affiliation{Department of Physics, University of Texas at Austin, Austin, TX 78712, USA}

\author{Wenbin Lu}
\affiliation{Department of Astronomy, University of Texas at Austin, Austin, TX 78712, USA}

\author{Pawan Kumar}
\affiliation{Department of Astronomy, University of Texas at Austin, Austin, TX 78712, USA}

\author{Rodolfo Santana}
\affiliation{Department of Astronomy, University of Texas at Austin, Austin, TX 78712, USA}





\begin{abstract}
We study the spectra of photospheric emission from highly relativistic gamma-ray burst outflows using a Monte Carlo (MC) code. We consider the Comptonization of photons with a fast cooled synchrotron spectrum in a relativistic jet with realistic photon to electron number ratio $N_{\gamma}/N_e = 10^5$, using mono-energetic protons which interact with thermalised electrons through Coulomb interaction. The photons, electrons and protons are cooled adiabatically as the jet expands outwards. We find that the initial energy distribution of the protons and electrons do not have any appreciable effect on the photon peak energy $E_{\gamma,peak}$ and the power-law spectrum above $E_{\gamma,peak}$. The Coulomb interaction between the electrons and the protons does not affect the output photon spectrum significantly as the energy of the electrons is elevated only marginally. $E_{\gamma,peak}$ and the spectral indices for the low and high energy power-law tails of the photon spectrum remain practically unchanged even with electron-proton coupling. 
Increasing the initial optical depth $\tau_{in}$ results in slightly shallower photon spectrum below $E_{\gamma,peak}$ and fewer photons at the high-energy tail, although $f_{\nu} \propto \nu^{-0.5}$ above $E_{\gamma,peak}$ and up to $\sim 1$ MeV, independent of $\tau_{in}$. We find that $E_{\gamma,peak}$ determines the peak energy and the shape of the output photon spectrum. Lastly, we find that our simulation results are quite sensitive to $N_{\gamma}/N_e$, for $N_{e} = 3\times10^3$. 
For almost all our simulations, we obtain an output photon spectrum with power-law tail
above $E_{\gamma,peak}$ extending up to $\sim 1$ MeV.
\end{abstract}


\section{Introduction}
\label{Intro}

The radiation mechanism responsible for the prompt emission of Gamma-Ray Bursts (GRBs) is still not very well understood. The observed spectra is generally modelled using the Band function (\citealt{Band}), which is a smoothly connected broken power-law with observed peak energy $E_p \sim 300$ keV and non-thermal power-laws below and above the peak (in some cases up to $\sim$ GeV energies) (\citealt{Kaneko} and \citealt{Preece}). The two most widely explored models to explain the GRB spectrum are the internal dissipation model and the photospheric model (\citealt{RM94,Piran04,KZ15}).

In the internal dissipation model, the energy is dissipated either by internal shocks (\citealt{RM94}) or by magnetic reconnection in a Poynting dominated jet (\citealt{ZY11}). The prompt radiation is due to the synchrotron emission from non-thermal electrons gyrating in a shock-generated magnetic field (\citealt{Meszaros94,Piran99,LP00}). This model suffers from low radiation efficiency because only the kinetic energy associated with the differential motion of the shells can be dissipated and not the kinetic energy associated with the bulk motion of the jet (\citealt{Koba97,Lazzati99,K99,Guetta01,Kino04}). However, the observations confirm high efficiencies up to few tens of percent (\citealt{Zhang07}). Moreover, synchrotron emission cannot explain hard GRB spectra at low energies (\citealt{Preece98,Ghirlanda03}) and the spectrum is directly related to the radiation mechanism involved rather than interaction with baryons and leptons in the jet.

Owing to these shortcomings of the internal dissipation model, many researchers have recently focussed on the photospheric emission model (\citealt{MR00,RM05,LB10,Ito13,CL15,Santana16}). Unlike the internal dissipation model, the photospheric model can explain the observed high radiation efficiencies. The shape of the spectrum is determined by the interaction of photons with matter in the jet, which is through Compton scattering, and hence should be independent of the emission mechanism. There have been many successful attempts to explain the high-frequency non-thermal tails using sub-photospheric dissipation (\citealt{Pe'er06,Gia06,LB10,Vurm11,Ito13,CL15,Santana16}), however non-thermal tails at low energies still cannot be obtained (\citealt{PR11,CL15}). 

In this paper, we study the Comptonisation of seed photons produced by synchrotron emission of fast cooling electrons below the photosphere (\citealt{Ghisellini00,Granot00}). The electrons and protons are accelerated to relativistic energies by a dissipation mechanism such as internal shocks (\citealt{LB10,Toma11,Lazzati13}) or magnetic reconnection (\citealt{Thompson94,Gia06,Gia12}) at an optical depth of a few or larger. The choice of synchotron spectrum over thermal spectrum for photons is justified as there are not enough scatterings at relatively small optical depths $\tau \lesssim$ 10 to thermalise the photon spectrum (\citealt{Begue13}). Most of the energy in the jet is carried by the protons because of their large mass and the average energy of the electrons is assumed to be much larger as compared to the average energy of the photons. We consider sub-photospheric heating of electrons which occurs as a result of continuous injection of energy from the protons through the Coulomb interaction and is more physically motivated than episodic injection of energy (\citealt{Gia06,Pe'er06,LB10,Santana16}). Photons undergo multiple scatterings with the electrons and gain energy until the outflow becomes optically thin and the photons escape the photosphere. Unlike many previous photospheric MC simulations (\citealt{LB10,CL15}), we also include adiabatic cooling of electrons, protons and photons due to the expansion of the relativistic jet (\citealt{Santana16}). 

Almost all photospheric MC simulations performed previously used relatively small photon to electron ratio $N_{\gamma}/N_{e} \sim 10^{1} - 10^{4}$ (\citealt{LB10,CL15}), which leads to unrealistically low radiation efficiencies contradicting GRB observations (\citealt{Zhang07,Santana16}). In this work, we use $N_{\gamma}/N_{e} = 10^{5}$ which gives radiative efficiency $\eta \sim 10\%$ (consistent with observations) in addition to incorporating electron heating in a more realistic way to determine if the high-energy GRB prompt emission spectral index can be reproduced using the photospheric emission model. For all our simulations, we use Maxwell-Boltzmann (MB) electrons and mono-energetic protons as the respective seed distributions.

This paper is organized as follows. In Section \ref{Implementation}, we describe the physics and implementation of our MC photospheric code. We present our simulation results in Section \ref{Results} and discuss the interpretation of these results in Section \ref{Discussions}. Finally, we present our conclusions in Section \ref{Conclusions}.

\section{Implementation of the photospheric code}
\label{Implementation}
In this section, we describe the implementation of our MC code and give an overview of the basic physics included. We discuss how the energy and velocity distributions of the electrons, protons and photons are initialised and how they are affected by adiabatic cooling, Coulomb interaction and scattering events. The scattering events between the electrons and photons are performed one at a time in our MC code. Throughout this paper, primed quantities are in the jet-comoving frame while unprimed quantities are in the lab frame.

\subsection{Input parameters}
Here we describe the input parameters used for our MC simulations. 

\begin{itemize}

\item Isotropic equivalent luminosity of the jet, $L$ : We consider $L = 10^{52}$ ergs/sec for all our simulations (\citealt{Liang07,WP10}).
\item Bulk Lorentz factor of the jet, $\Gamma$ : For all our simulations, we consider $\Gamma =$ 300 (\citealt{Xue09,Liang10}).
\item Number of electrons in a simulation, $N_{e}$: Like in the previous photospheric simulations (\citealt{LB10,CL15,Santana16}), we consider $N_{e} = 10^3$. In Figure \ref{fig1}, we show that it is enough to use $10^3$ electrons for accurately simulating the GRB jet.
\item Number of photons in a simulation, $N_{\gamma}$ : We consider $N_{\gamma} = 10^8$ for our simulations (\citealt{Santana16}). This was done to ensure that $N_{\gamma}/N_{e} = 10^{5}$.
\item Number of protons in a simulation, $N_{p}$ : We consider $N_{p} = 10^3$ as $N_{e} = N_{p}$ due to charge neutrality of the jet.
\item Number of photons collected for the output spectrum, $N_{\gamma,\rm{collect}}$ : Like in the previous simulations (\citealt{LB10,Santana16}), we consider $N_{\gamma,\rm{collect}} = N_{\gamma}/3$ as it gives us a time-averaged representation of the GRB spectrum by allowing for enough photon-electron scatterings to accurately represent the output spectrum.
\item Initial optical depth, $\tau_{in}$ : The initial optical depth determines the distance from the central engine where all the electrons, photons and protons are injected. We consider $\tau_{in} = 2, 4, 8$ and 16 in this work.
\item Seed photon spectrum : We consider the synchrotron spectrum for fast cooling electrons where the energy distribution is given by smoothly connected power-laws (\citealt{Granot00,Piran04}):
\begin{equation}
\label{Eqn1}
f_{\nu} = \left\{
\begin{array}{ll}
\left(\frac{\nu_{ac}}{\nu_{sa}}\right)^{11/8}\:\left(\frac{\nu}{\nu_{ac}}\right)^{2}, & \nu_{l} < \nu < \nu_{ac}\\
\left(\frac{\nu}{\nu_{sa}}\right)^{11/8}, & \nu_{ac} < \nu < \nu_{sa} \\
\left(\frac{\nu}{\nu_{sa}}\right)^{-1/2}, & \nu_{sa} < \nu < \nu_{m} \\
\left(\frac{\nu_{m}}{\nu_{sa}}\right)^{-1/2}\:\left(\frac{\nu}{\nu_{m}}\right)^{-p/2}, & \nu_{m} < \nu < \nu_{u}\\
\end{array}
\right. 
\end{equation}
where $f_{\nu}$ is the flux per unit frequency in the lab frame.
Throughout this paper we consider, $h\nu_{l}^{\prime} = 3\times10^{-9}$ eV, $h\nu_{ac}^{\prime} = 2\times10^{-2}$ eV, $h\nu_{sa}^{\prime} = 2$ eV, $h\nu_{m}^{\prime} = 1$ keV and $h\nu_{u}^{\prime} = 30$ keV, which is justified by the choice of our parameters and the typical values of other parameters: $\epsilon_{B} = 0.1$, $\epsilon_{e} = 0.1$, $N = 10^{2}$ and $T = 10$ s (\citealt{Granot00}). $f_{\nu}$ is peak normalised and the high energy spectral index $p = 2.5$ (\citealt{KZ15}). 
\item Electron distribution : We consider Maxwell-Boltzmann (MB) distribution of electrons with the initial $\gamma_{e,in}^{\prime}$ as the input parameter. For our simulations, $\gamma_{e,in}^{\prime} = 25, 50, 75 \rm{\:and \:} 100$. 
\item Proton distribution : For our simulations, we consider mono-energetic distribution of protons with the initial $\gamma_{p,in}^{\prime}$ as the input parameter. We perform the simulations using $\gamma_{p,in}^{\prime} = 1.5, 2, 5 \rm{\:and \:} 10$. 

\end{itemize}

\subsection{Initialisation of electrons, protons and photons}
At the beginning of our photospheric MC code, we initialise the directions and energies of all the electrons, protons and photons.

\subsubsection{Direction and energy of electrons and protons}
The initial directions of the velocities of $N_{e}$ electrons and $N_{p}$ protons are chosen randomly in the comoving frame of the jet (see Appendix B1 of \citealt{Santana16}). For the initial energies of the electrons, $\gamma_{e}^{\prime}$ is chosen from the relativistic MB distribution corresponding to temperature $T_{e,in}^{\prime}$ which is given by (see Appendix B2.1 of \citealt{Santana16}),
\begin{equation}
\label{Eqn2}
k_{B}T_{e,in}^{\prime} = (\gamma_{ad,e,in}^{\prime} - 1)(\gamma_{e,in}^{\prime} - 1)m_{e}c^{2}
\end{equation}
where, the electron adiabatic index $\gamma_{ad,e,in}^{\prime} \approx (4\gamma_{e,in}^{\prime} + 1)/(3\gamma_{e,in}^{\prime})$. For the mono-energetic protons, $\gamma_{p}^{\prime} = \gamma_{p,in}^{\prime}$. We assume that initially all the $N_{e}$ electrons and $N_{p}$ protons are distributed uniformly in the comoving frame of the jet.

\subsubsection{Direction, energy and position of photons}
The initial directions of the velocities of $N_{\gamma}$ photons are chosen randomly in the comoving frame of the jet (see Appendix C1 of \citealt{Santana16}). The initial energies of the photons in the comoving frame of the jet is chosen from the synchrotron radiation distribution of fast cooling electrons as given in Equation \ref{Eqn1} (see Appendix \ref{AppendixA} for algorithm).

The position of the $N_{\gamma}$ photons are assigned randomly and they are uniformly distributed within a cone with solid angle $1/\Gamma$ pointing towards the observer. The initial distance from the central engine (in the lab frame) where the photons are injected is given by
\begin{equation}
\label{Eqn3}
R_{in} = \frac{L\sigma_{T}}{8\pi m_{p}c^{3}\beta \Gamma^{3}\tau_{in}}
\end{equation}
where $\beta = \sqrt{1 - (1/\Gamma^2)}$ and $\sigma_{T}$ is the Thomson cross section.

\subsection{Adiabatic cooling of electrons, protons and photons}
The energies of the electrons, protons and photons decreases as the jet expands outward, due to adiabatic cooling. Due to the adiabatic cooling, the energy of the electrons [protons] decreases by a factor $R^{-2(\gamma_{ad,e}^{\prime} - 1)}$ [$R^{-2(\gamma_{ad,p}^{\prime} - 1)}$] where $\gamma_{ad,e}^{\prime} \approx (4\gamma_{e}^{\prime}+1)/(3\gamma_{e}^{\prime})$ [$\gamma_{ad,p}^{\prime} \approx (4\gamma_{p}^{\prime}+1)/(3\gamma_{p}^{\prime})$] is the adiabatic index of the electron [proton] and $R$ is the radial distance the jet has travelled from the central engine. For the photons, the drop in energy is by a factor $R^{-2/3}$. These expressions are valid because the electron density $n_{e}^{\prime}$ drops by a factor $R^{2}$ as the relativistic outflow expands outward and the radial width of the jet remains unchanged. After each scattering event, the energies of the electrons, protons and photons are modified due to adiabatic cooling as
\begin{equation}
\label{Eqn4}
\frac{\gamma_{e,f}^{\prime} - 1}{\gamma_{e,i}^{\prime} - 1} = \left(\frac{R_{in} + (t_{\gamma} + \Delta t_{\gamma})\beta c}{R_{in} + t_{e}\beta c}\right)^{-2(\gamma_{ad,e,i}^{\prime} - 1)}
\end{equation}
\begin{equation}
\label{Eqn5}
\frac{\gamma_{p,f}^{\prime} - 1}{\gamma_{p,i}^{\prime} - 1} = \left(\frac{R_{in} + (t_{\gamma} + \Delta t_{\gamma})\beta c}{R_{in} + t_{e}\beta c}\right)^{-2(\gamma_{ad,p,i}^{\prime} - 1)}
\end{equation}
\begin{equation}
\label{Eqn6}
\frac{E_{\gamma,f}^{\prime}}{E_{\gamma,i}^{\prime}} = \left(\frac{R_{in} + (t_{\gamma} + \Delta t_{\gamma})\beta c}{R_{in} + t_{\gamma}\beta c}\right)^{-2/3}
\end{equation}
where $R_{in}$ is given by Equation \ref{Eqn3}. The subscripts $i$ and $f$ are used to denote the energies before and after the photon has travelled a distance $s^{\prime}$ in the comoving frame of the jet. The total time elapsed in the lab frame for the photon and electron (which undergo scattering) is given by $t_{\gamma}$ and $t_{e}$, respectively. The time needed by the photon to travel a distance $s^{\prime}$ in the lab frame is given by $\Delta t_{\gamma}$ (see Appendix C3 of \citealt{Santana16} for Lorentz transformation). The proton is considered to be moving with the electron and hence can be represented by the same time $t_{e}$ as it is practically unaffected by the photon-electron scattering event. After the photon travels a distance $s^{\prime}$, the electron and the photon reach the same final radial position where they interact by IC/Compton scattering.

\subsection{Coulomb interaction}
In addition to adiabatic cooling, the energies of the electrons and the protons are also affected by the Coulomb interaction between them. As the protons have much larger energies as compared to the electrons, electrons are always heated due to the energy transfer from the protons. Moreover, the electrons exchange energy between themselves and attain MB distribution after reaching equilibrium. Below we discuss how the electron and proton energies are affected due to these interactions.

\subsubsection{Electron-proton (e-p) interaction}
The timescale for Coulomb cooling of protons in the jet-comoving frame is (\citealt{Schlickeiser02}),
\begin{equation}
\label{Eqn7}
t_{p,Coul}^{\prime} = \frac{(\gamma_{p}^{\prime} - 1)m_{p}c^{2}}{5\times10^{-19} n_{e}^{\prime}} \frac{(8.3\times10^{-15} T_{e}^{\prime 3/2} + \beta_{p}^{\prime 3})}{\beta_{p}^{\prime 2}}
\end{equation}
where $n_{e}^{\prime}$ is the electron density in the jet-comoving frame, $T_{e}^{\prime}$ is the temperature of the electrons in the jet-comoving frame and $\beta_{p}^{\prime}$ is the speed of the protons divided by the speed of light.  The electron density $n_{e}^{\prime}$ is given by
\begin{equation}
\label{Eqn8}
n_{e}^{\prime} = \frac{L}{4\pi (R_{in} + t_{e}\beta c)^{2} m_{p}c^{3} \Gamma^{2}}
\end{equation}
The energies of the protons and electrons are modified due to Coulomb interaction after each scattering event.
The expressions used to update the $\gamma_{e}^{\prime}$ of an electron and $\gamma_{p}^{\prime}$ of a proton due to Coulomb interaction are
\begin{equation}
\label{Eqn9}
\gamma_{e,f}^{\prime} = \gamma_{e,i}^{\prime} + \frac{5\times10^{-19} n_{e}^{\prime}}{\Gamma m_{e} c^{2}} \frac{\beta_{p,i,avg}^{\prime 2} (t_{\gamma} + \Delta t_{\gamma} - t_{e})}{(8.3\times10^{-15} T_{e,i,avg}^{\prime 3/2} + \beta_{p,i,avg}^{\prime 3})}
\end{equation}
\begin{equation}
\label{Eqn10}
\gamma_{p,f}^{\prime} = \gamma_{p,i}^{\prime} - \frac{5\times10^{-19} n_{e}^{\prime}}{\Gamma m_{p} c^{2}} \frac{\beta_{p,i,avg}^{\prime 2} (t_{\gamma} + \Delta t_{\gamma} - t_{e})}{(8.3\times10^{-15} T_{e,i,avg}^{\prime 3/2} + \beta_{p,i,avg}^{\prime 3})}
\end{equation}
As before, the subscripts $i$ and $f$ are used to denote the energies before and after the photon travels by a distance $s^{\prime}$ in the jet-comoving frame. The factor of $1/\Gamma$ is included to transform the time from the lab frame to the jet-comoving frame. As the electrons experience Coulomb heating due to the average proton distribution around them and vice-versa, we include averaged quantities $\beta_{p,i,avg}^{\prime}$ and $T_{e,i,avg}^{\prime}$ which denote the speed of protons averaged over $N_{p}$ protons in the jet-comoving frame divided by the speed of light and the temperature of electrons corresponding to $\gamma_{e,i}^{\prime}$ averaged over $N_{e}$ electrons in the jet-comoving frame (see Equation \ref{Eqn2}), respectively. Thus, after each scattering event, the electrons gain some energy from the protons which is determined by their respective energy distributions.

\subsubsection{Electron-electron (e-e) interaction}
In addition to interacting with the protons, the electrons also exchange energy between themselves. The energy distribution of the electrons at thermal equilibrium is given by MB distribution with the peak temperature $T_{e,avg}^{\prime}$ determined by $\gamma_{e,avg}^{\prime}$ (see Equation \ref{Eqn2}). As the nature of the interaction between the electrons is the same as that with the protons, the timescale for this interaction can be obtained just by replacing the proton parameters with the electron parameters in Equation \ref{Eqn7},
\begin{equation}
\label{Eqn11}
t_{e,Coul}^{\prime} = \frac{(\gamma_{e,avg}^{\prime} - 1)m_{e}c^{2}}{5\times10^{-19} n_{e}^{\prime}} \frac{(8.3\times10^{-15} T_{e,avg}^{\prime 3/2} + \beta_{e,avg}^{\prime 3})}{\beta_{e,avg}^{\prime 2}}
\end{equation}
where $\beta_{e}^{\prime}$ is the speed of electron in the jet-comoving frame divided by the speed of light and all the electron parameters are averaged over all $N_{e}$ electrons. After each photon-electron scattering event, the average (over $N_{e}$ electrons) total time elapsed in the lab frame is evaluated for the electrons, which is denoted by $t_{e,avg}$. Whenever $t_{e,avg}$ exceeds any multiple of $t_{e,Coul} = \Gamma t_{e,Coul}^{\prime}$, the electron distribution is re-initialised to a MB distribution with $T_{e,avg}^{\prime}$ determined by $\gamma_{e,avg}^{\prime}$ at that point of the simulation.

It should be noted that the electron distribution in between consecutive scattering events can deviate from Maxwellian for large values of $N_{\gamma}/N_{e}$ (see Figure \ref{fig7}). In that case, the electron temperature $T_{e,avg}^{\prime}$ evaluated from Equation \ref{Eqn2} using $\gamma_{e,avg}^{\prime}$ may not exactly correspond to that of a Maxwellian with the same energy. However, Equations \ref{Eqn7} and \ref{Eqn11} can still be used to model the Coulomb interactions fairly well as long as: (1) the quasi-Maxwellian distribution is unimodal with peak energy close to that of the approximated Maxwellian distribution, and (2) the timescale at which the electrons are re-initialised to Maxwellian distribution is comparable to the electron-photon scattering timescale. Both these conditions are satisfied for all our simulations and the electron distribution need not be updated after every scattering event which is computationally very expensive.

\begin{figure*}
\gridline{\fig{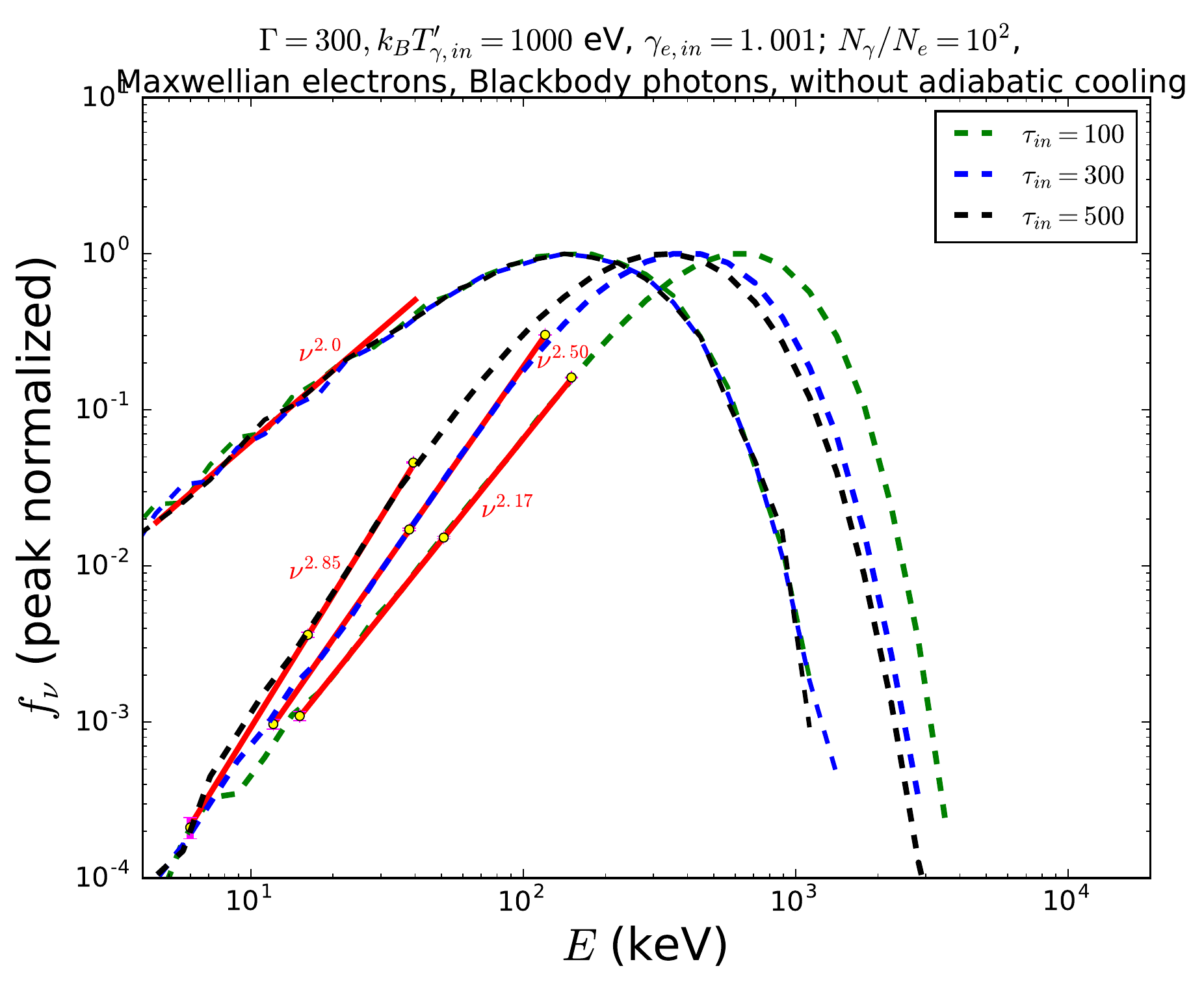}{0.5\textwidth}{}
          \fig{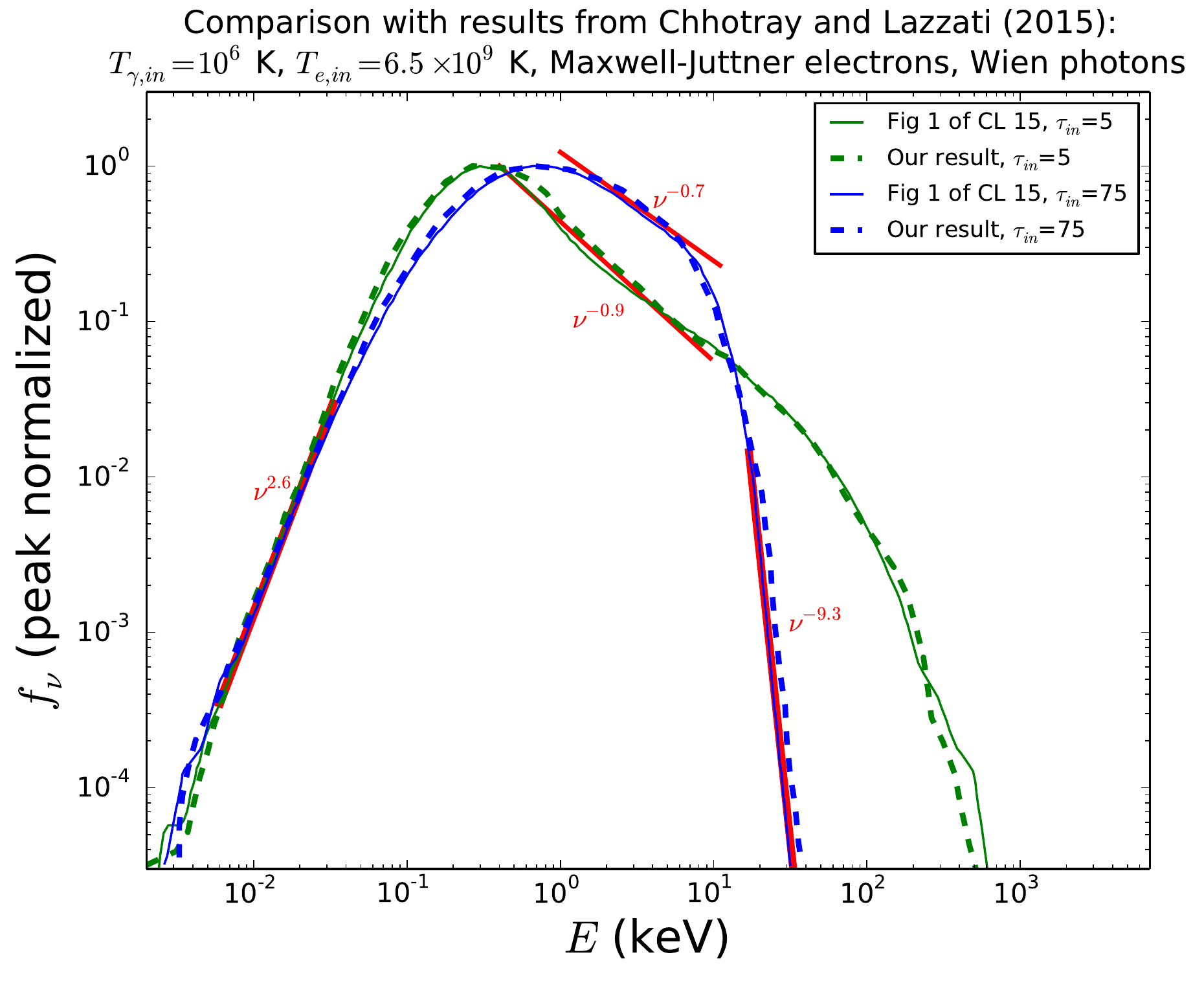}{0.5\textwidth}{}
          } \vspace{-2em} 
  \caption{Code validation tests.
  {\it Left Panel:} Obtaining equilibrium electron and photon distributions for $\Gamma=300$, $k_{B}T_{\gamma,in}^{\prime} = 1000$ eV, $N_{\gamma}/N_{e} = 2\times10^{6}/2\times10^{4}$ and no adiabatic cooling for Maxwellian electrons with constant $\gamma_{e,in}^{\prime} = 1.001$ and $\tau_{in} = 100$, 300 and 500. The low-energy spectral indices are: $\alpha_{\tau_{in}} = 2.17\pm0.07$, $2.50\pm0.08$ and $2.85\pm0.20$ for $\tau_{in} = 100$, 300 and 500 respectively. 
The error bars for few selected points (yellow dots) on the photon distribution are shown - most of the error bars are too small to see except for low energies where the Poisson fluctuations are considerable due to small photon numbers.
The electron and photon peak energies do not coincide but differ by a factor of $\sim 2$ as their average energies are different by a factor of 2 for the same equilibrium temperature. 
  {\it Right Panel:} Comparison of our simulation results (dashed lines) with those from Figure 1 of \citealt{CL15} (solid lines) for Wien photons with $T_{\gamma,in} = 10^6$ K, $N_{\gamma}/N_{e} = 10^{3}$ and no adiabatic cooling for Maxwell-Juttner electrons with $T_{e,in} = 6.5\times10^9$ K and $\tau_{in} = 5$ and 75.
  }
  \label{fig0}
\end{figure*}

\begin{figure*}         
\gridline{\fig{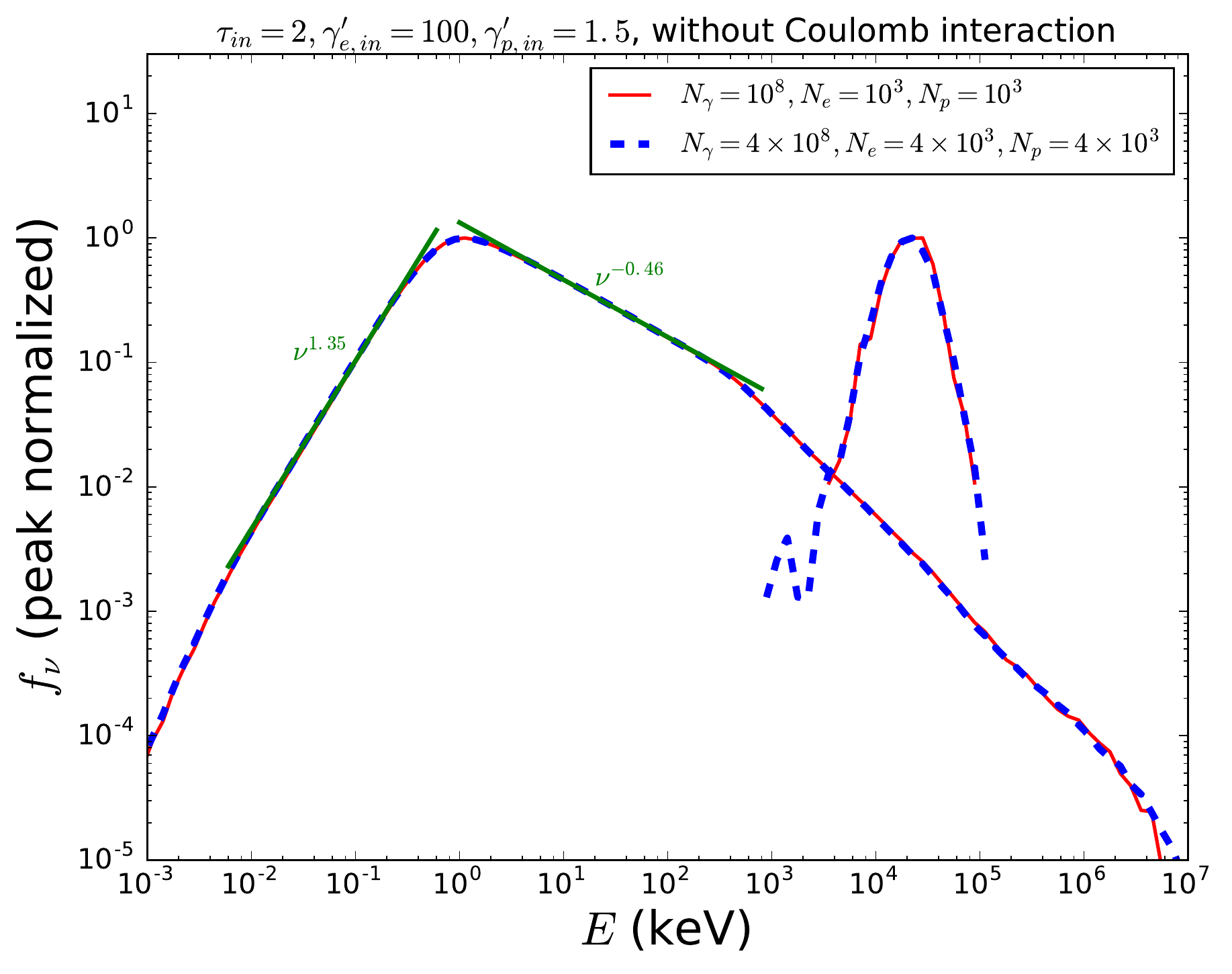}{0.5\textwidth}{}
          \fig{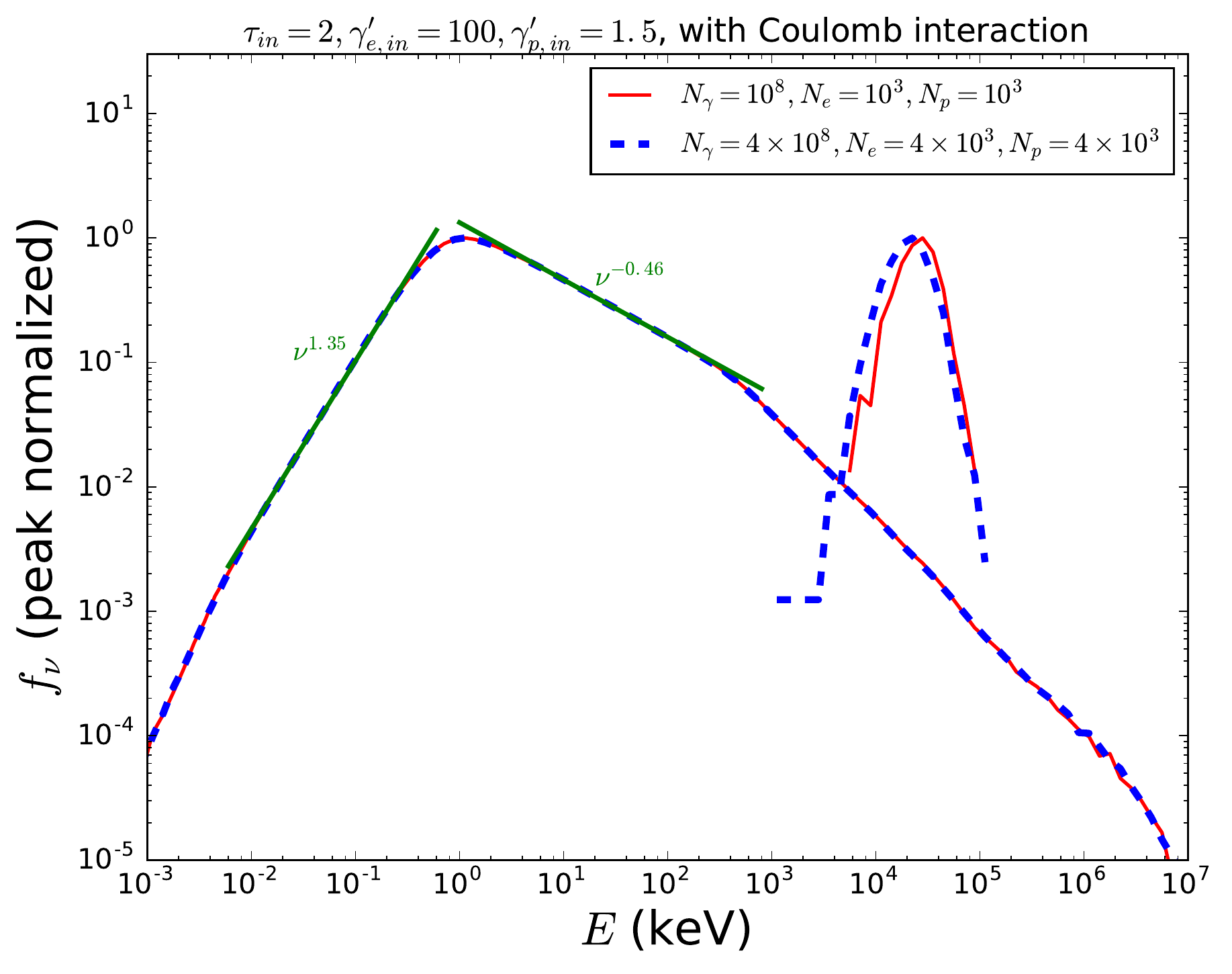}{0.5\textwidth}{}
          }  \vspace{-2em}
  \caption{Comparison of $N_{\gamma} = 10^{8}, N_{e} = 10^{3}, N_{p} = 10^{3}$ simulation results with $N_{\gamma} = 4\times10^{8}, N_{e} = 4\times10^{3}, N_{p} = 4\times10^{3}$ simulation results for photons with seed spectrum given by Equation \ref{Eqn1}, $\gamma_{e,in}^{\prime} = 100$ and $\gamma_{p,in}^{\prime} = 1.5$ for $\tau_{in} = 2$. 
 {\it Left Panel:} When only adiabatic cooling is considered.
 {\it Right Panel:} When both adiabatic cooling and Coulomb interaction (e-p and e-e) are considered.
 }
  \label{fig1} 
\end{figure*}

\subsection{Main photospheric code}
\label{Sec2.5}
At the beginning of the simulation, the distance $s^{\prime}$ that each photon travels in the comoving frame of the jet before scattering an electron is drawn randomly using the formula $s^{\prime} = - l_{mfp}^{\prime}\rm{ln}(\alpha)$ (\citealt{Santana16}). Here, $l_{mfp}^{\prime} = 1/(n_{e}^{\prime}\sigma_{T})$ is the mean free path of the photons in the jet-comoving frame and $\alpha$ is a uniformly distributed random number within 0 and 1. Once $s^{\prime}$ for all $N_{\gamma}$ photons are drawn, the photons are propagated and their new positions are Lorentz transformed to the lab frame (see Appendix C3 of \citealt{Santana16}) and compared with the photospheric distance $R_{ph}$ ($R$ corresponding to $\tau = 1$ in Equation \ref{Eqn3}) to check if any photon escapes the photosphere without interacting with an electron. For the photons which escape the photosphere, the energies are Doppler boosted to the lab frame and are stored. All other photons are placed in a priority queue $(t_{\gamma,l},l)$, where $t_{\gamma,l}$ denotes the total time elapsed in the lab frame for the photon with index $l$. The photon properties such as position, direction and energy can be accessed using the respective photon index $l$. The priority queue structure allows for the photon with the smallest $t_{\gamma,l}$ to get scattered first (\citealt{Santana16}).

Next, we propagate the first photon in the priority queue using the corresponding $s^{\prime}$. One of the $N_{p}$ protons is chosen randomly while one of the $N_{e}$ electrons is selected by sampling the electron-photon scattering probability distribution function (see Appendix \ref{AppendixB} for algorithm) given by,

\begin{equation}
\label{Eqn12}
P_{scatt}(\beta_{e}^{\prime},\theta_{e}^{\prime}) = \frac{1}{4\pi \beta_{e}^{\prime 2}}(1 - \beta_{e}^{\prime}\: \rm{cos}\theta_{e}^{\prime})
\end{equation}
where, $\theta_{e}^{\prime}$ is the angle between the electron and photon directions before the scattering event in the jet-comoving frame and $\beta_{e}^{\prime}$ is the speed of the electron in the jet-comoving frame divided by the speed of light.

Next, Equations \ref{Eqn4} - \ref{Eqn6} are used to update the energies after adiabatic cooling and Equations \ref{Eqn9} - \ref{Eqn10} are used to update the energies after Coulomb interaction. Once the energies are determined, the dimensionless photon energy in the electron rest frame $z_{i}$ and the scattering cross-section $\sigma(z_{i}^{\prime})$ are calculated (see Appendix D of \citealt{Santana16}). A uniformly distributed random number $0 \le \alpha_{s} \le 1$ is drawn and is compared with the scattering probability $\sigma(z_{i}^{\prime})/\sigma_{T}$ to determine whether the electron-photon scattering event actually happens. If $\alpha_{s} \le \sigma(z_{i}^{\prime})/\sigma_{T}$ is satisfied, the scattering event takes place and the direction and energy of the photon is updated along with the direction and energy of the electron (see Appendices D and E of \citealt{Santana16}).

\begin{figure*}
\gridline{\fig{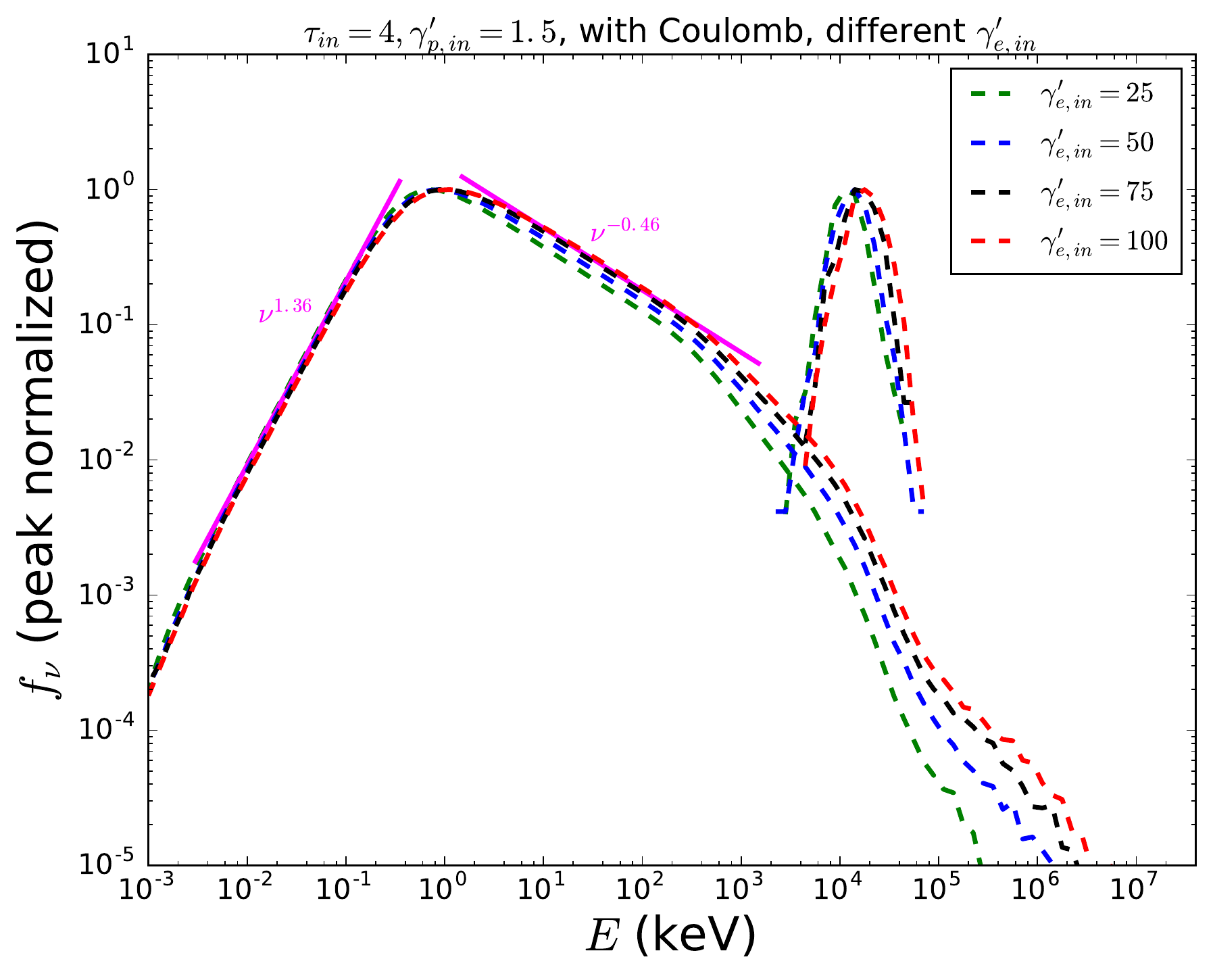}{0.5\textwidth}{}
          \fig{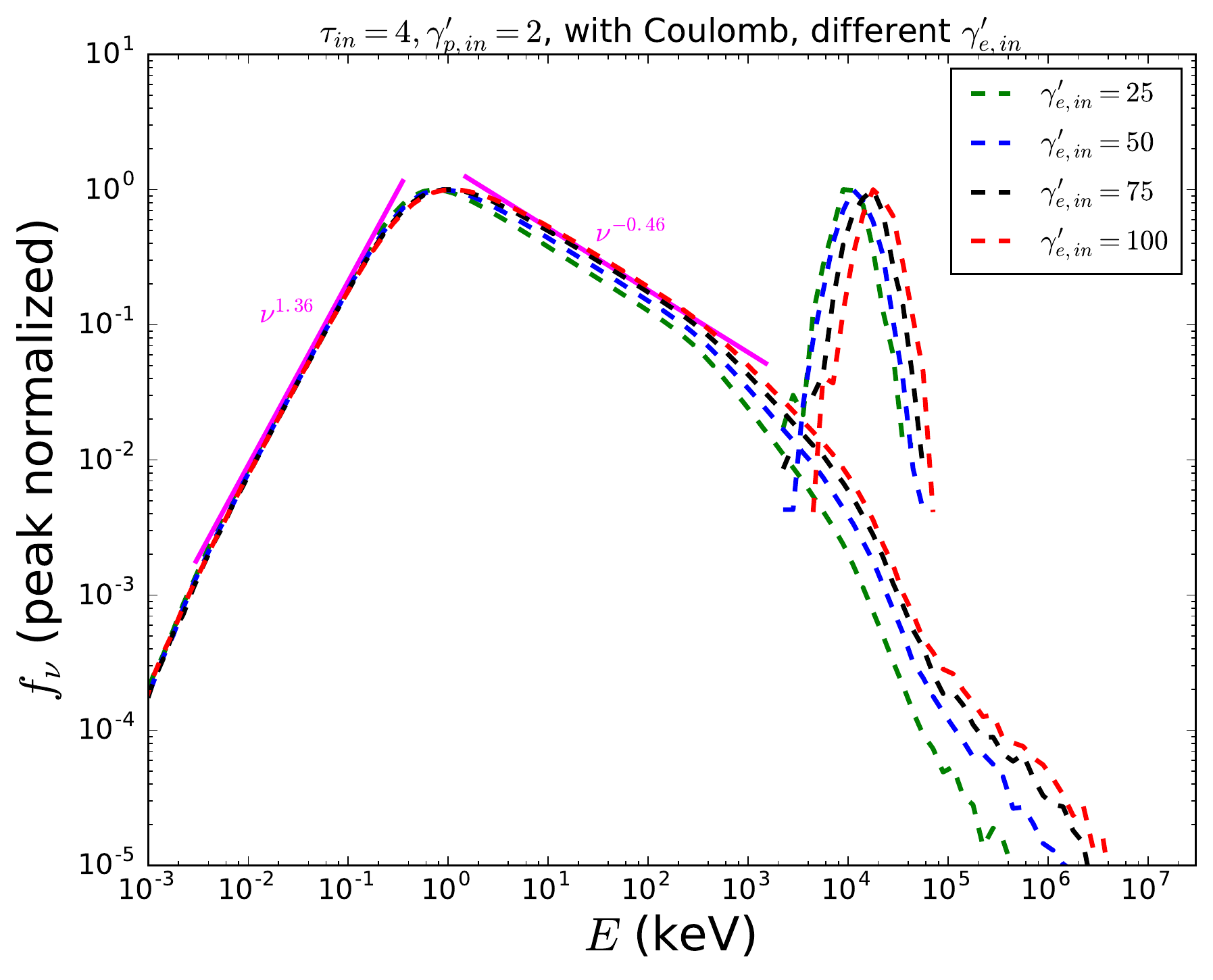}{0.5\textwidth}{}
          }  \vspace{-2em}        
\gridline{\fig{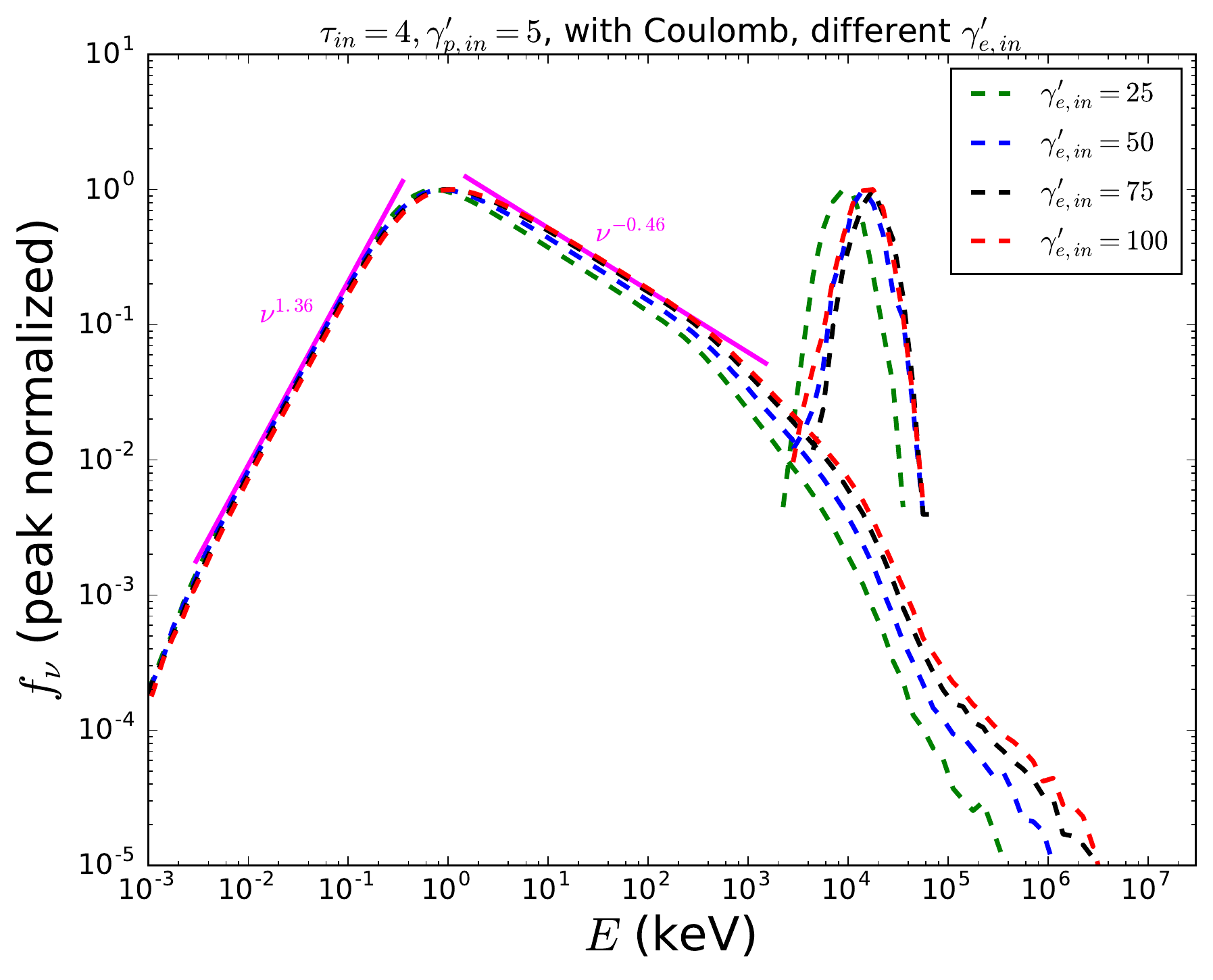}{0.5\textwidth}{}
          \fig{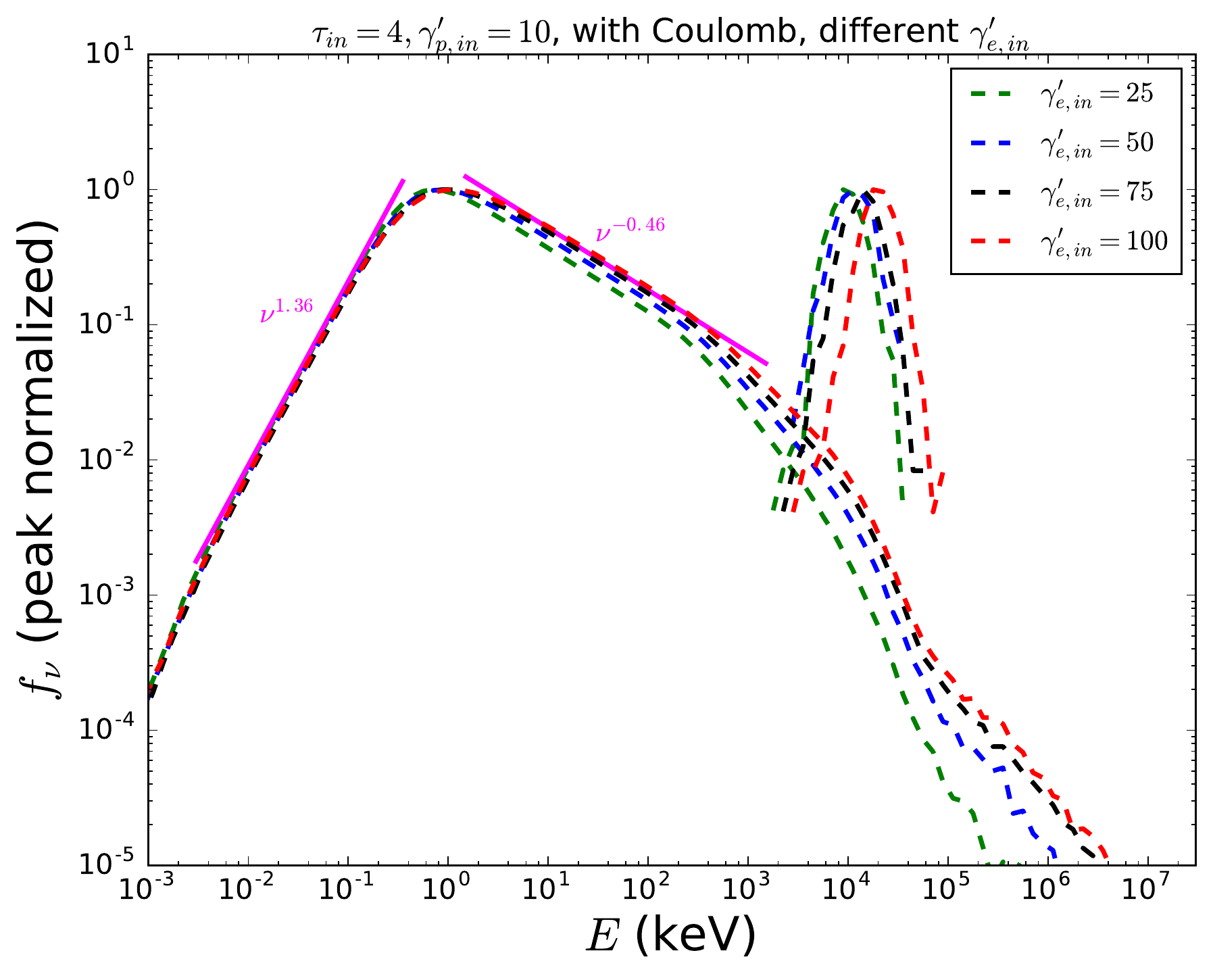}{0.5\textwidth}{}
          }  \vspace{-2em}        
  \caption{Simulation results for photons with seed spectrum given by Equation \ref{Eqn1}, $\tau_{in} = 4$ and with Coulomb interaction (both e-p and e-e). 
 {\it Top-Left Panel:} For $\gamma_{p,in}^{\prime} = 1.5$ and $\gamma_{e,in}^{\prime}$ = 25, 50, 75 and 100. 
 {\it Top-Right Panel:} For $\gamma_{p,in}^{\prime} = 2$ and $\gamma_{e,in}^{\prime}$ = 25, 50, 75 and 100. 
 {\it Bottom-Left Panel:} For $\gamma_{p,in}^{\prime} = 5$ and $\gamma_{e,in}^{\prime}$ = 25, 50, 75 and 100.
 {\it Bottom-Right Panel:} For $\gamma_{p,in}^{\prime} = 10$ and $\gamma_{e,in}^{\prime}$ = 25, 50, 75 and 100. }
  \label{fig2} 
\end{figure*}


Finally, irrespective of whether the photon is scattered or not, a new $s^{\prime}$ is drawn at its current location and the photon is propagated as was done at the beginning of the simulation. The distance travelled by the photon is Lorentz transformed to the lab frame and whether the current location of the photon $R$ exceeds $R_{ph}$ is checked. If $R \ge R_{ph}$, the energy of the photon is Doppler boosted to the lab frame and is stored. Else, the photon is again placed in the priority queue with the updated total elapsed time in the lab frame $t_{\gamma,l}$ and the whole process is repeated until $N_{\gamma,collect}$ photons escape the photosphere.

\subsection{Photospheric code tests}
We first try to reproduce the equilibrium distributions for electrons and photons undergoing Compton scatterings to check the validity of our simulations. In the left panel of Figure \ref{fig0}, we present the results of simulations in which Maxwellian electrons are held fixed at energy $\gamma_{e,in}^{\prime}=1.001$ while they scatter Blackbody photons with $k_{B}T_{\gamma,in}^{\prime} = 1000$ eV for $N_{\gamma}/N_{e}=10^{2}$ and $\Gamma=300$. The electrons and the photons are not cooled due to adiabatic expansion of the jet as the initial optical depth is varied, $\tau_{in}=100$, 300 and 500. For photons interacting with electrons kept at a constant temperature bath, the equilibrium distribution at large $\tau_{in}$ approaches Bose-Einstein distribution with a non-zero chemical potential while the electrons attain Maxwell-Boltzmann distribution. 

We find that the equilibrium distribution (within uncertainty) for photons ($f_{\nu} \propto \nu^{3}$ at low energies and $f_{\nu} \propto e^{-\nu}$ at high energies) and electrons ($f_{\nu} \propto \nu^{2}$ at low energies and $f_{\nu} \propto e^{-\nu}$ at high energies) is obtained close to $\tau_{in} \sim 500$. It should be noted that the spectral indices obtained from the power-law fits in the left panel of Figure \ref{fig0} (and the rest of the figures in this paper) have statistical uncertainties due to the non-zero energy bin width, $\Delta E_{bw} =1$ eV, and the relatively small number of photons at low energies (for $N_{\gamma}=10^8$) in our simulation. However, these uncertainties are typically very small for the parameters that we consider and can be ignored. 

In the right panel of Figure \ref{fig0}, we compare our simulation results with that of Figure 1 of \cite{CL15} for the same input parameters: Wien photons with $T_{\gamma,in} = 10^6$ K, $N_{\gamma}/N_{e} = 10^{3}$ and no adiabatic cooling for Maxwell-Juttner electrons with $T_{e,in} = 6.5\times10^9$ K and $\tau_{in} = 5$ and 75. We find that there is good agreement of our results with \citet{CL15} for both the simulations which demonstrates that our photospheric code is working properly. 

We then check whether $N_{e} = 10^{3}$, $N_{p} = 10^{3}$ and $N_{\gamma} = 10^{8}$ are appropriate choices for representing the electron, proton and photon distributions in the relativistic jet. MC photospheric simulations have been performed previously with $N_{e} = 10^{3}$ (\citealt{LB10,CL15}), but smaller $N_{\gamma}/N_{e} \sim 10^{1} - 10^{4}$ were considered for those simulations. Simulations have also been performed with thermal photons as the seed spectrum to show that $N_{e} = 10^3$ is enough to represent the electron distribution for $N_{\gamma}/N_{e} = 10^{5}$ (\citealt{Santana16}), although e-p and e-e interactions were neglected in those simulations.
We perform simulations with $N_{\gamma} = 10^{8}$, $N_{e} = 10^{3}$ and $N_{p} = 10^{3}$ and compare them with $N_{\gamma} = 4\times10^{8}$, $N_{e} = 4\times10^{3}$ and $N_{p} = 4\times10^{3}$ in Figure \ref{fig1}. For both panels, $\tau_{in} = 2$, $\gamma_{e,in}^{\prime} = 100$ and $\gamma_{p,in}^{\prime} = 1.5$ are considered and the seed photon distribution is given by Equation \ref{Eqn1}. The left panel shows simulations performed without considering Coulomb interactions whereas the right panel shows simulations where both e-p and e-e Coulomb interactions were considered. The very good agreement between the simulation results suggests that $N_{e} = N_{p} = 10^{3}$ is enough for accurately representing relativistic jets for $N_{\gamma}/N_{e} = 10^{5}$.

\section{Simulation results}
\label{Results}

In this section, we present the results of our photospheric MC simulations. In all the figures, the photon energy spectrum and the electron kinetic energy spectrum is in the lab frame (Doppler boosted from the jet-comoving frame by multiplying with $\Gamma$) at the end of each simulation. Unless stated otherwise, e-p and e-e interactions are considered for the simulations.

In Figure \ref{fig2}, we present the simulation results for different combinations of $\gamma_{e,in}^{\prime}$ (=25, 50, 75, 100) and $\gamma_{p,in}^{\prime}$ (=1.5, 2, 5, 10) at $\tau_{in} = 4$ when e-p and e-e interactions are considered. Comparing the different panels, we can see that $\gamma_{p,in}^{\prime}$ does not have any effect on the output spectra. 
However, the electrons are more energetic at the end of the simulation for larger $\gamma_{e,in}^{\prime}$: $\gamma_{e}^{\prime} \sim 1.065$ for $\gamma_{e,in}^{\prime} = 25$ and $\gamma_{e}^{\prime} \sim 1.130$ for $\gamma_{e,in}^{\prime} = 100$.
Unlike previous simulations for $N_{\gamma}/N_{e} = 10^{5}$ (\citealt{Santana16}), our output photon spectrum does not have a sharp drop in $f_{\nu}$ after $E_{\gamma,peak}$. The photon spectra show a power-law after the peak, $f_{\nu} \propto \nu^{-0.5}$ upto $\sim 10^3$ keV. After $\sim 10^{4}$ keV, $f_{\nu}$ for photons drops sharply by $\sim 2$ orders of magnitude. This is due to the fact that the average electron energy $\Gamma (\gamma_{e,avg}^{\prime} - 1) m_e c^2$ is $\sim 10^{4}$ keV, beyond which enough photons cannot be upscattered by the electrons. 
The photon spectrum extends to higher energies for larger $\gamma_{e,in}^{\prime}$ ($\sim 4\times10^{5}$ keV for $\gamma_{e,in}^{\prime} = 25$ to $\sim 4\times10^{6}$ keV for $\gamma_{e,in}^{\prime} = 100$) as the highest energy that a photon with energy $E_{\gamma,peak}^{\prime}$ can get upscattered to after one scattering is $\sim E_{\gamma,peak}^{\prime}\Gamma \gamma_{e,in}^{\prime 2}$.
More energetic electrons (with larger $\gamma_{e,in}^{\prime}$) can transfer more energy to the photons which results in higher $f_{\nu}$ at large energies $\sim 10^{4} - 10^{7}$ keV.

\begin{figure*}
\gridline{\fig{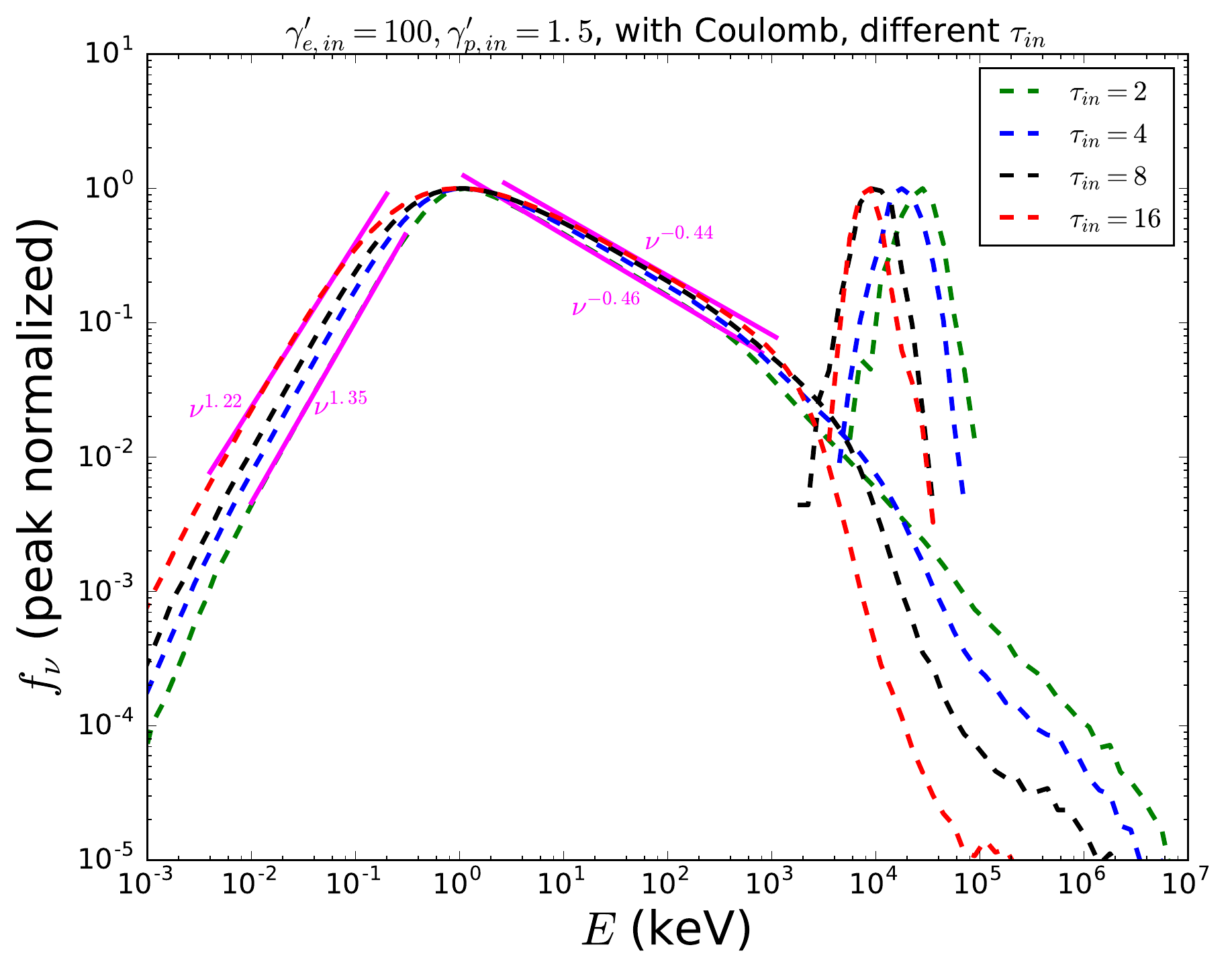}{0.5\textwidth}{}
          \fig{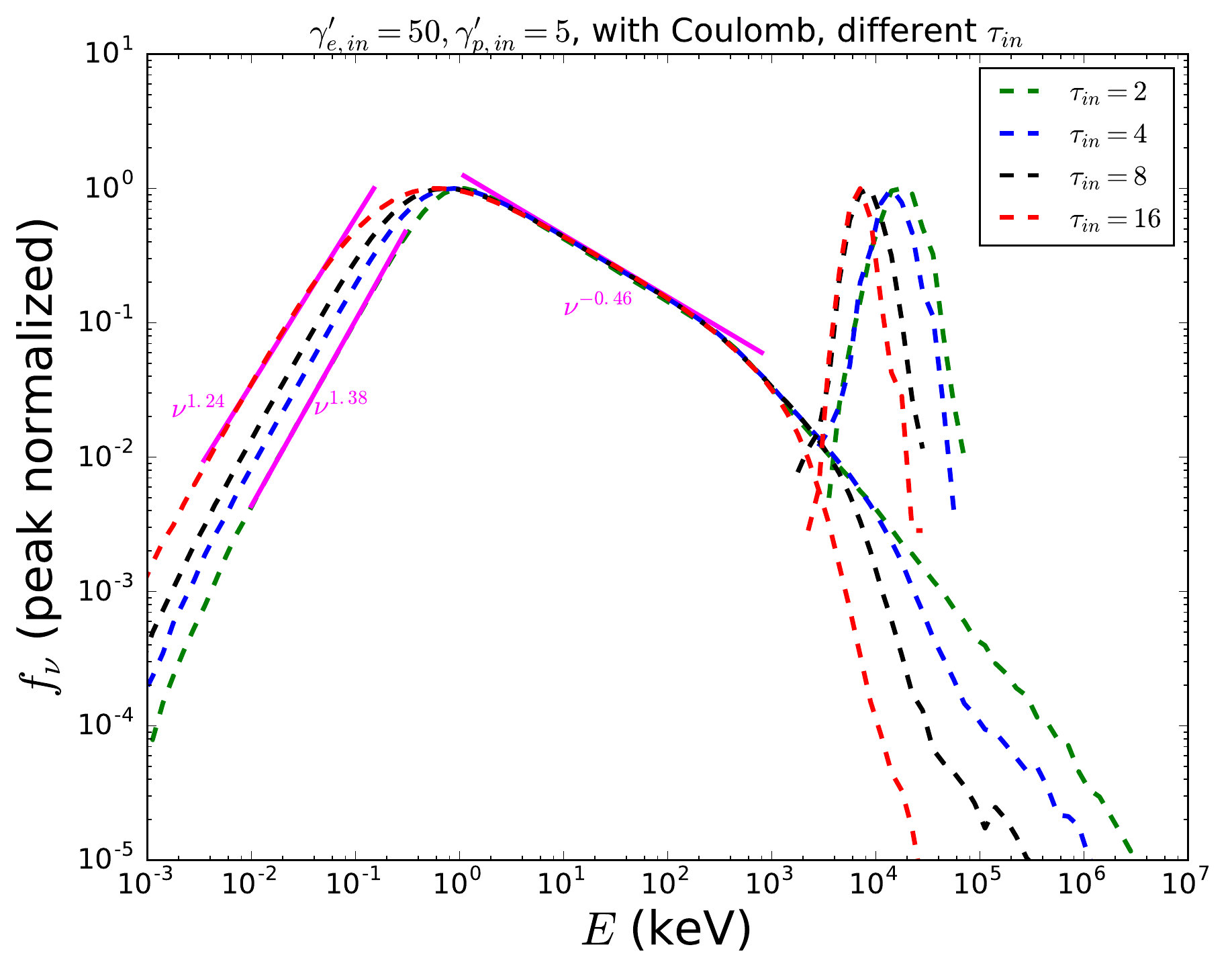}{0.5\textwidth}{}
          }  \vspace{-2em}
  \caption{Comparison of simulation results for photons with seed spectrum given by Equation \ref{Eqn1}, with Coulomb interaction (e-p and e-e) and different $\tau_{in} = 2, 4, 8$ and 16. 
 {\it Left Panel:} For $\gamma_{p,in}^{\prime} = 1.5$ and $\gamma_{e,in}^{\prime}$ = 100. 
 {\it Right Panel:} For $\gamma_{p,in}^{\prime} = 5$ and $\gamma_{e,in}^{\prime}$ = 50.
 }
  \label{fig4} 
\end{figure*}

\begin{figure*}
\gridline{\fig{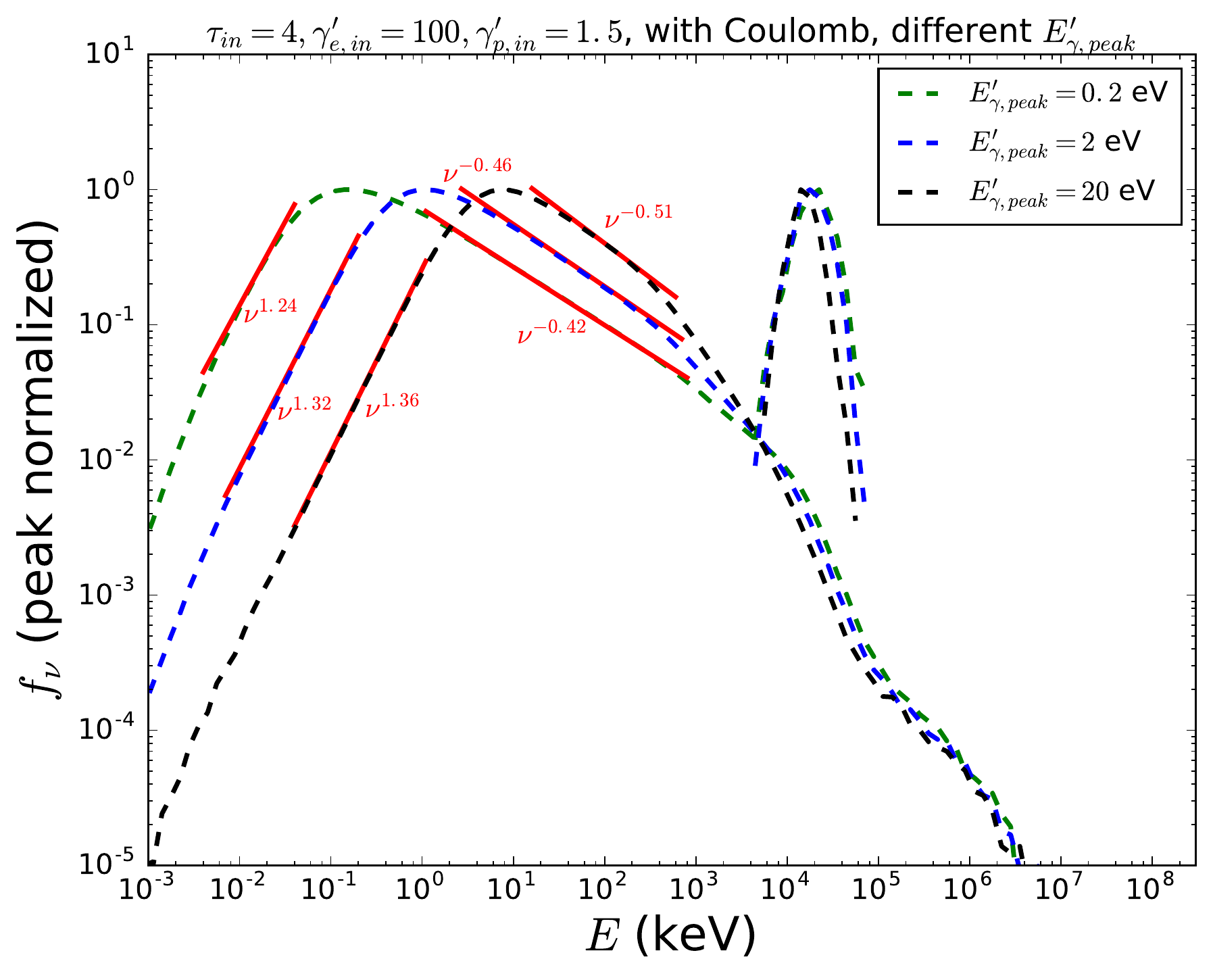}{0.5\textwidth}{}
          \fig{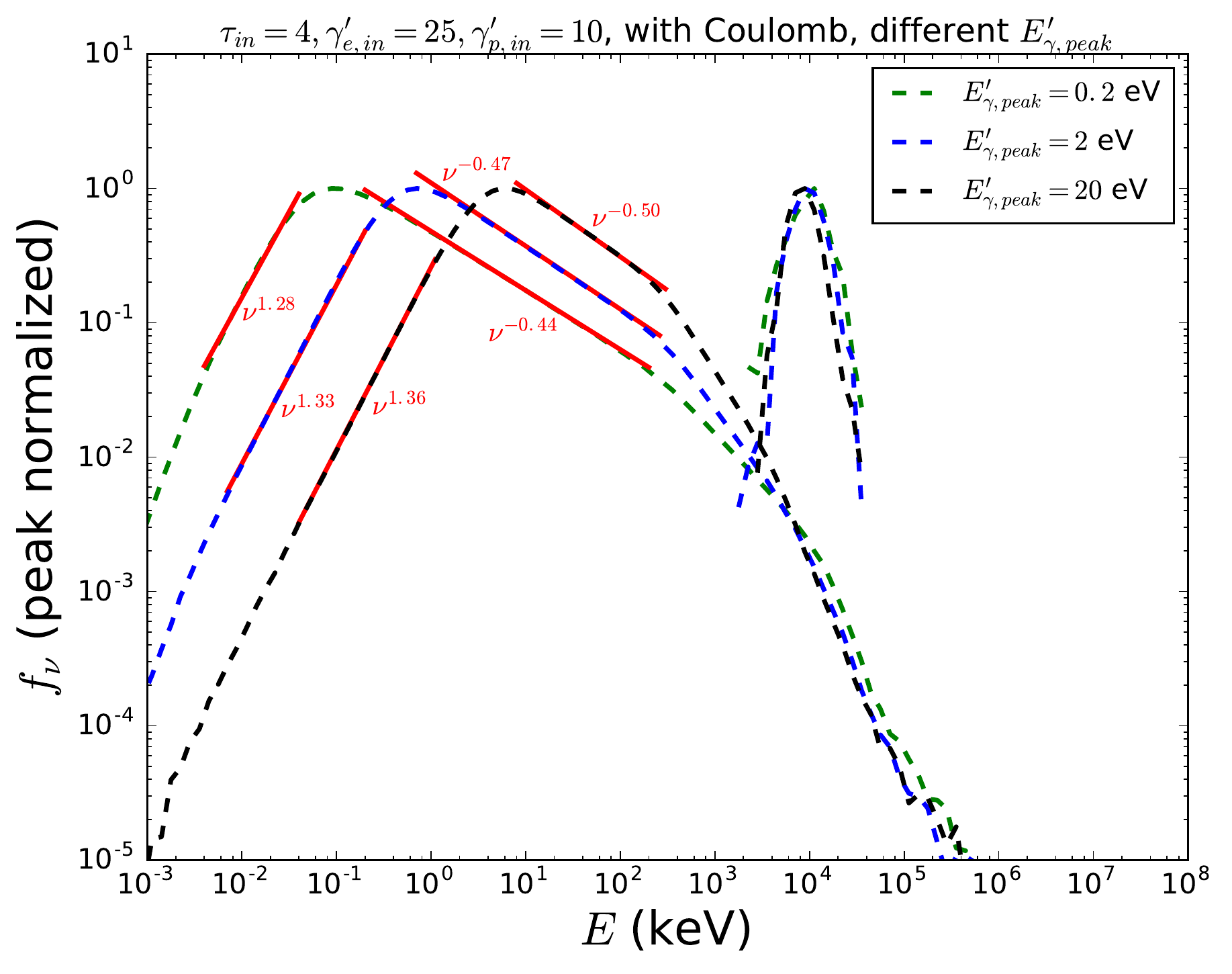}{0.5\textwidth}{}
          }  \vspace{-2em}
 \caption{Comparison of simulation results for photons with seed spectrum given by Equation \ref{Eqn1} and different $E_{\gamma,peak}^{\prime} = h\nu_{sa}^{\prime}$ = 0.2 eV, 2 eV and 20 eV, for $\tau_{in}$ = 4 and with Coulomb interaction (e-p and e-e). 
 {\it Left Panel:} For $\gamma_{p,in}^{\prime} = 1.5$ and $\gamma_{e,in}^{\prime}$ = 100. 
 {\it Right Panel:} For $\gamma_{p,in}^{\prime} = 10$ and $\gamma_{e,in}^{\prime}$ = 25.}
  \label{fig5} 
\end{figure*}

In Figure \ref{fig4}, we present the simulation results for two different combinations of $\gamma_{e,in}^{\prime}$ and $\gamma_{p,in}^{\prime}$ for $\tau_{in} =$ 2, 4, 8 and 16 when Coulomb (e-p and e-e) interaction is considered. As $\tau_{in}$ increases, the peak energy of the electron and the photon output spectrum shifts to lower energies which is due to adiabatic cooling (see Equations \ref{Eqn4} and \ref{Eqn6}). 
The energy of the electrons at the end of the simulation drops from $\gamma_{e}^{\prime} \sim 1.196$ ($\gamma_{e}^{\prime} \sim 1.130$) for $\tau_{in}=2$ to $\gamma_{e}^{\prime} \sim 1.052$ ($\gamma_{e}^{\prime} \sim 1.046$) for $\tau_{in}=16$ when $\gamma_{p,in}^{\prime} = 1.5$ and $\gamma_{e,in}^{\prime} = 100$ ($\gamma_{p,in}^{\prime} = 5$ and $\gamma_{e,in}^{\prime} = 50$). 
While there are smaller number of photons at higher energies for larger $\tau_{in}$, the photon spectrum becomes slightly shallower at energies below $E_{\gamma,peak}$. As in the previous case considered, the photon spectrum shows a power-law $f_{\nu} \propto \nu^{-0.5}$ right after the peak and upto $\sim 10^{3}$ keV even though $\tau_{in}$ changes considerably. 
The output photon spectrum becomes shallower below $E_{\gamma,peak}$ for larger $\tau_{in}$: $f_{\nu} \propto \nu^{1.4}$ for $\tau_{in} = 2$ to $f_{\nu} \propto \nu^{1.2}$ for $\tau_{in} = 16$.
Multiple scatterings become more probable with increasing $\tau_{in}$ which results in photons getting scattered different number of times by the electrons before escaping out of the photosphere (\citealt{Poz83}). 
As a result, the output photon spectrum broadens and becomes shallower below $E_{\gamma,peak}$ (see Figure \ref{fig8}).

\begin{figure*}
\gridline{\fig{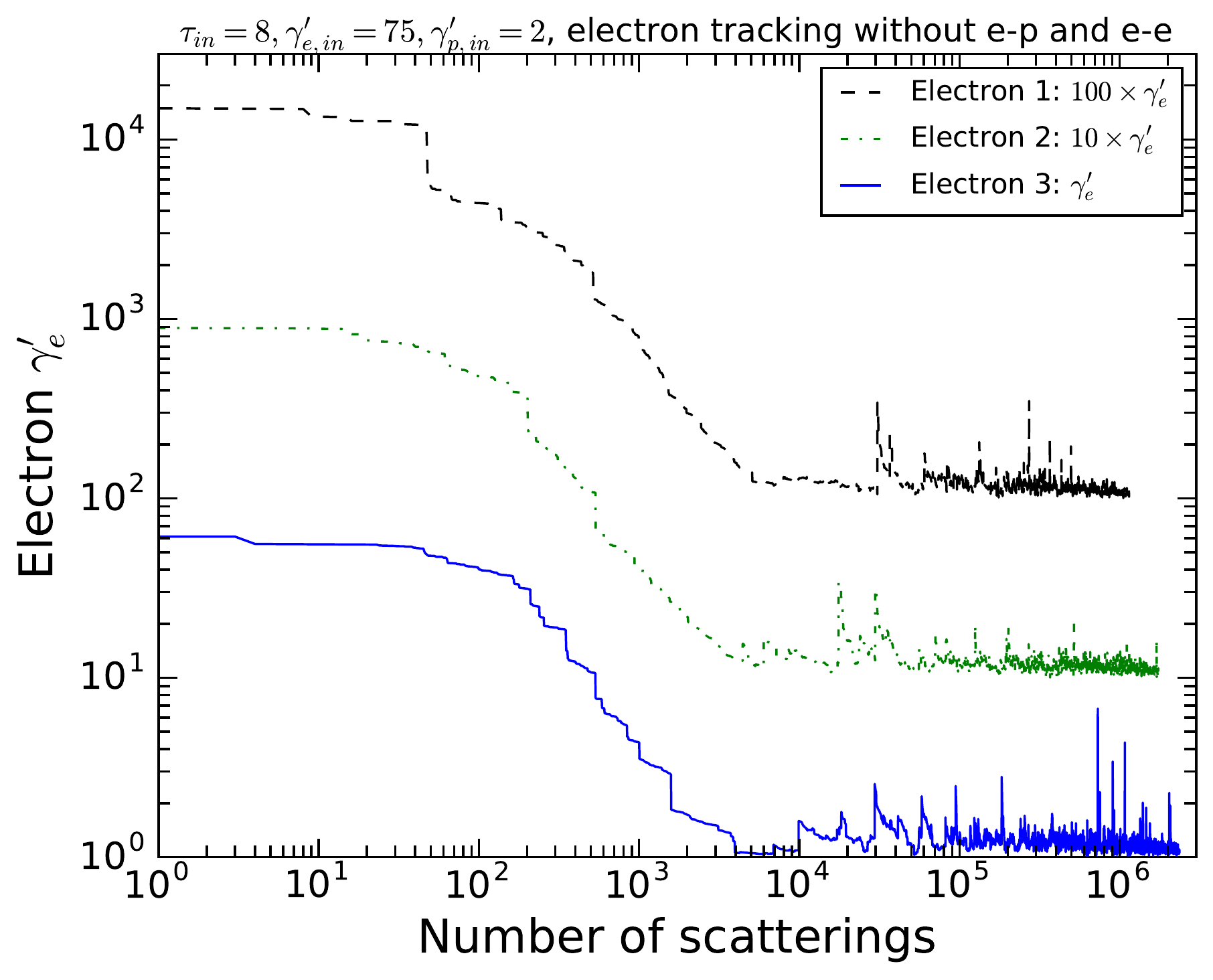}{0.5\textwidth}{}
          \fig{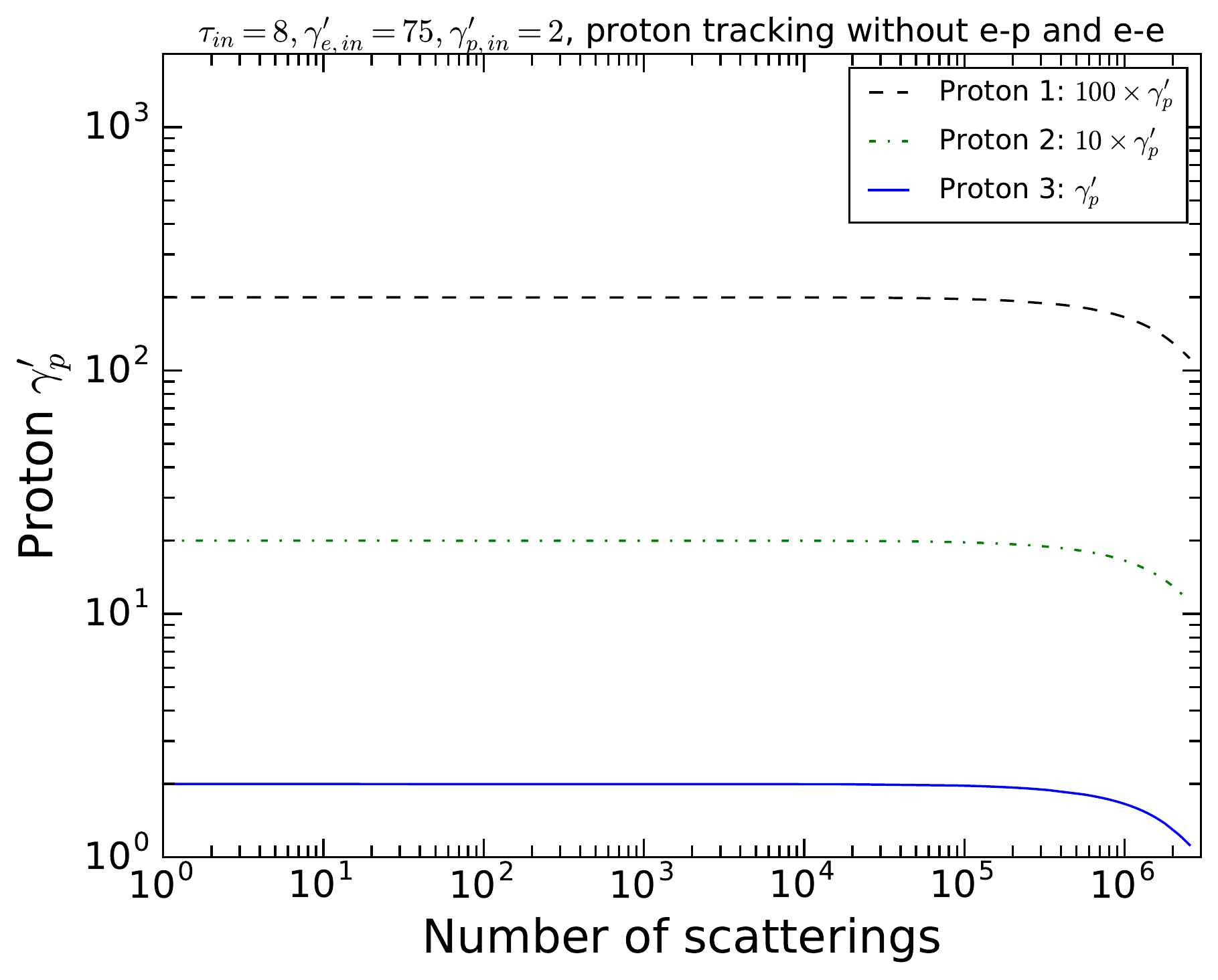}{0.5\textwidth}{}
          }      \vspace{-2em}   
\gridline{\fig{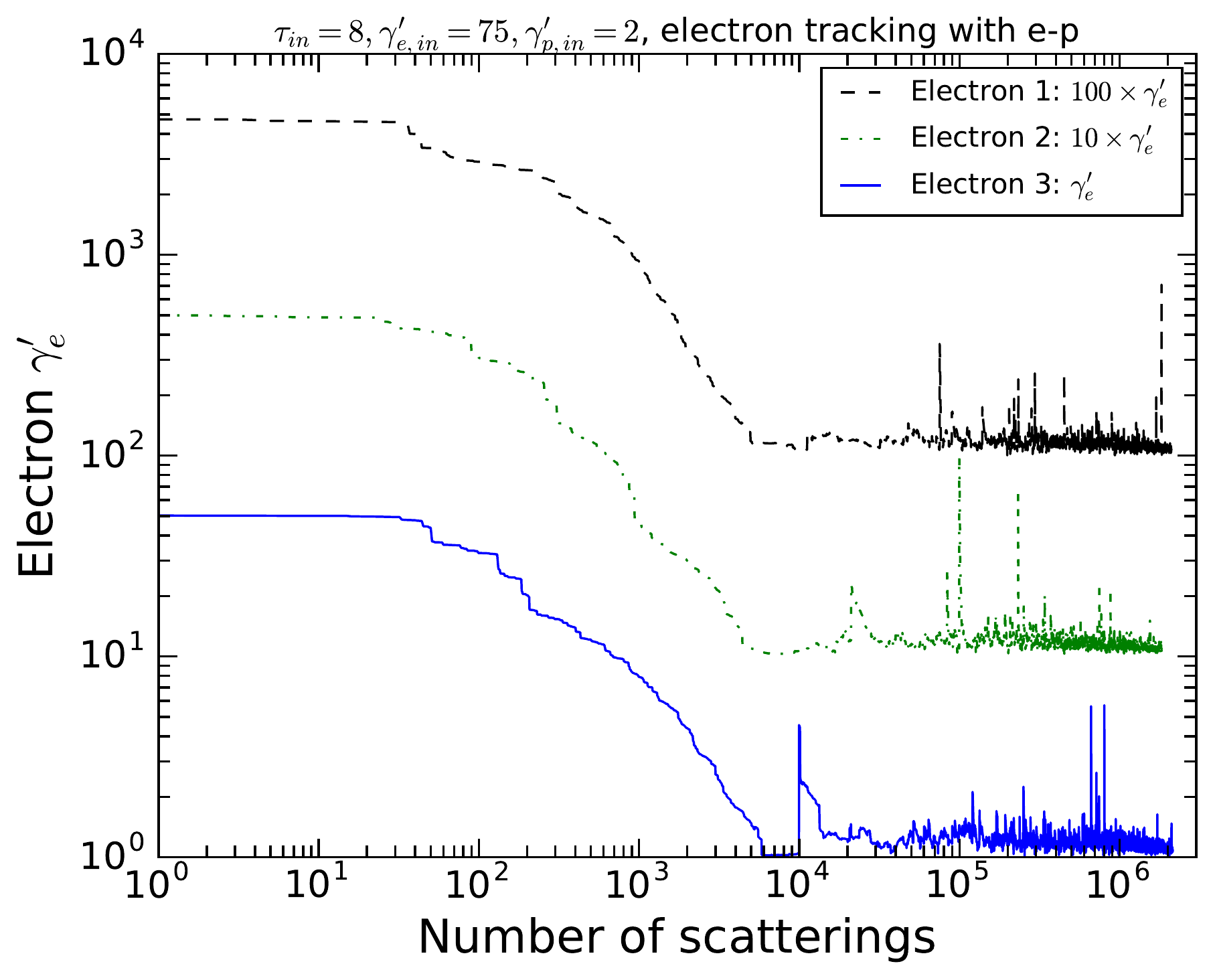}{0.5\textwidth}{}
          \fig{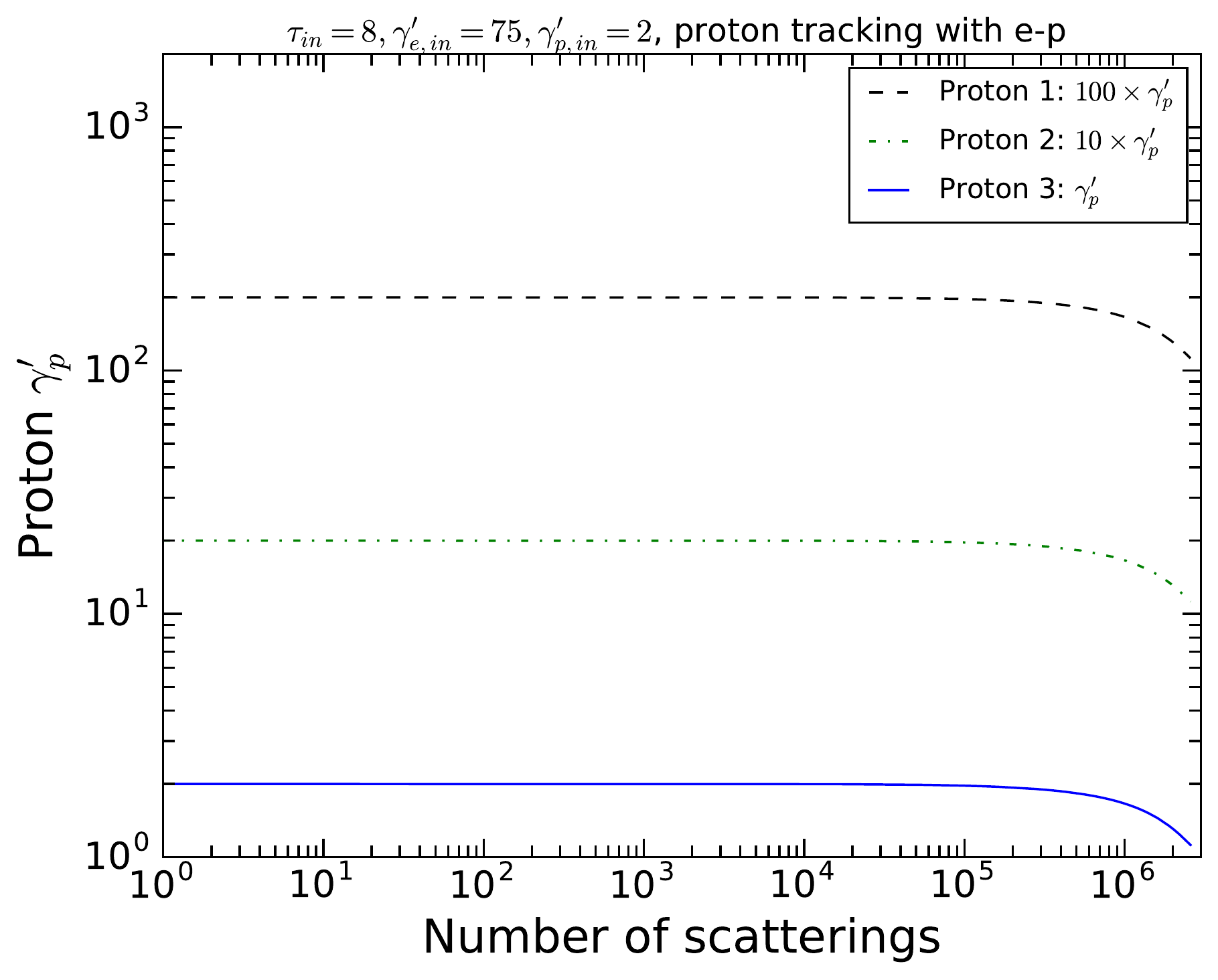}{0.5\textwidth}{}
          }      \vspace{-2em}    
\gridline{\fig{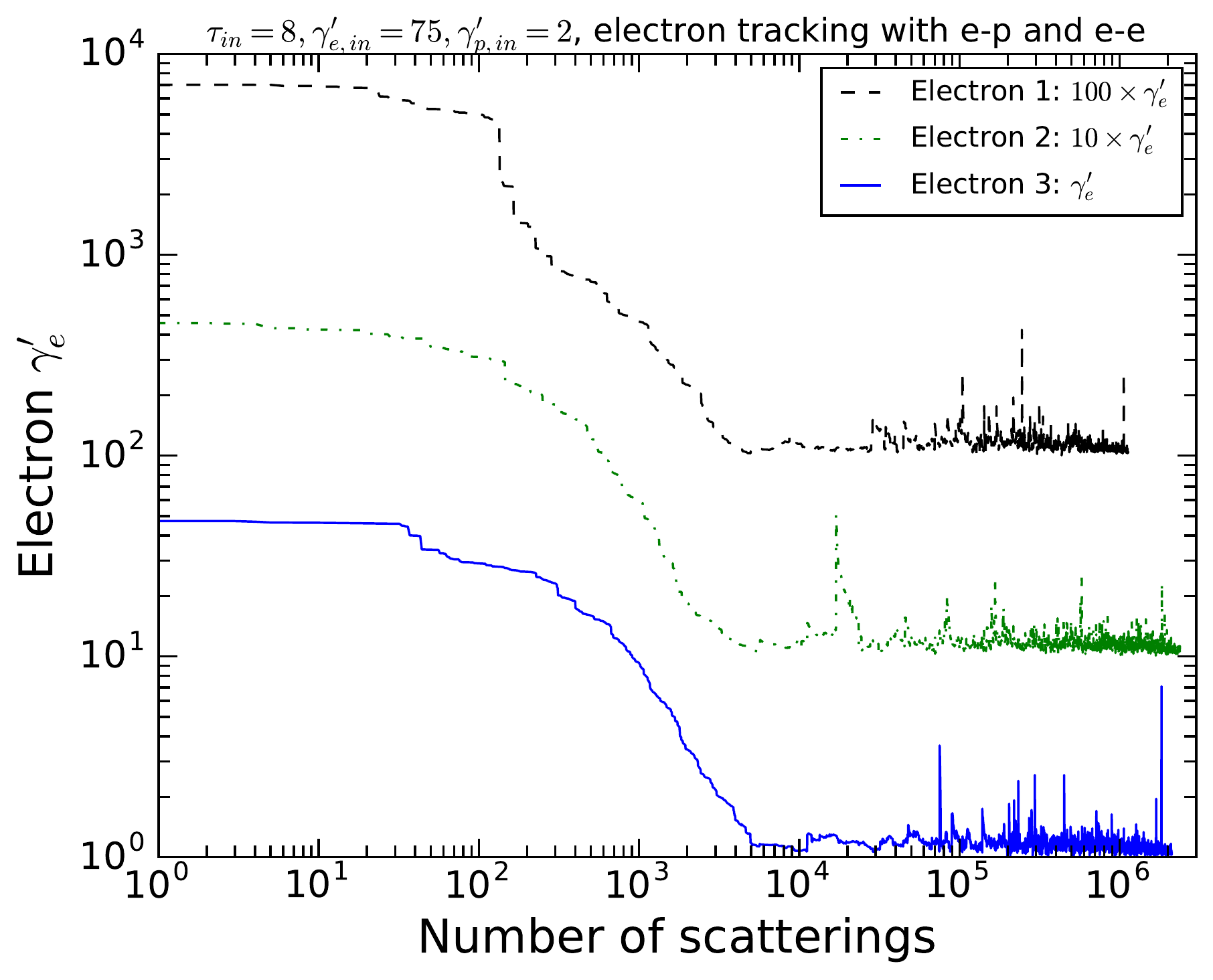}{0.5\textwidth}{}
	\fig{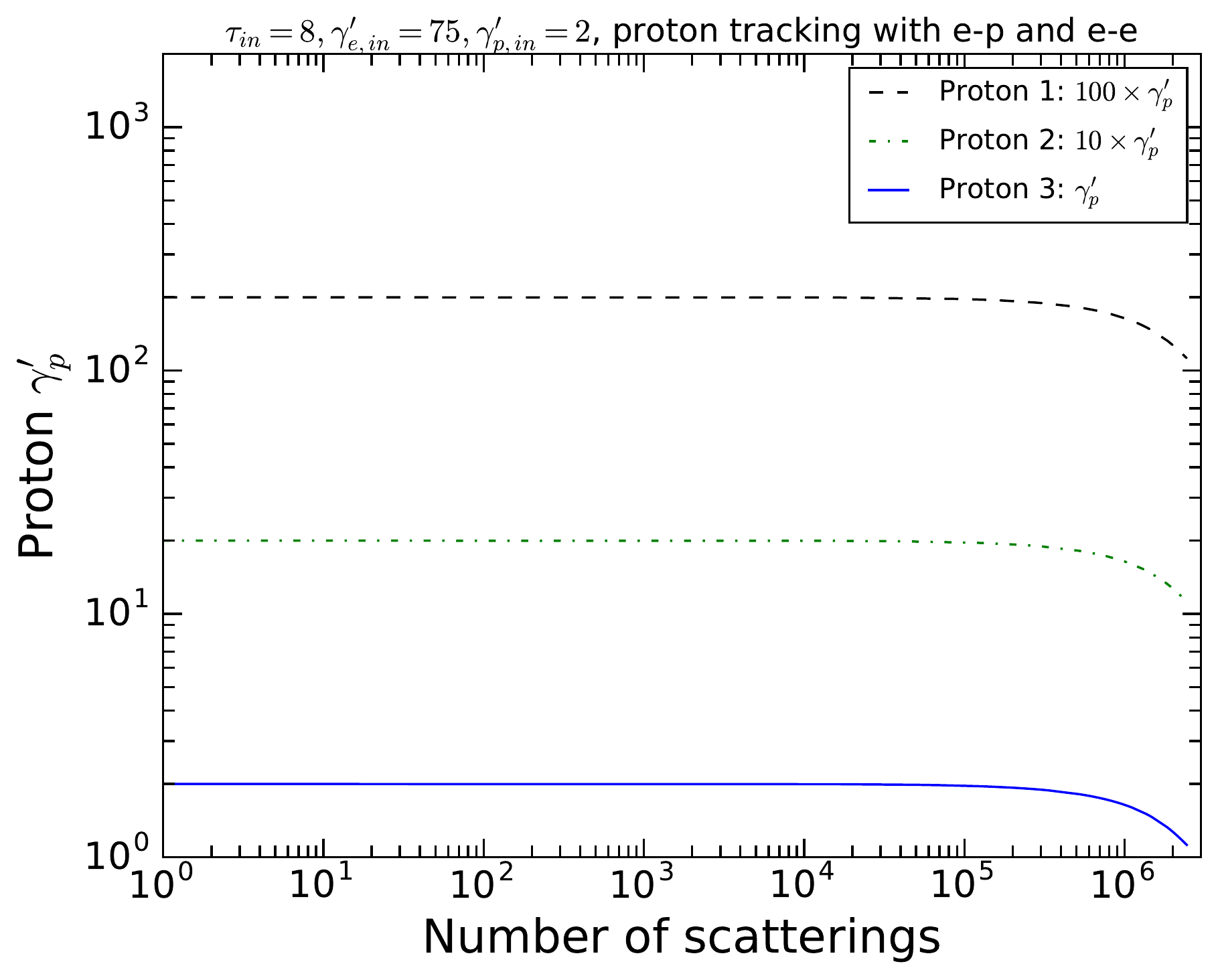}{0.5\textwidth}{}
	}      \vspace{-2em}
  \caption{Evolution of $\gamma_{e}^{\prime}$ for 3 electrons and $\gamma_{p}^{\prime}$ for 3 protons, for photons with seed spectrum given by Equation \ref{Eqn1} and $\tau_{in} = 8$, $\gamma_{e,in}^{\prime}$ = 75 and $\gamma_{p,in}^{\prime}$ = 2. 
  {\it Top-Left and Top-Right Panels:} Without e-p and e-e interactions.
  {\it Middle-Left and Middle-Right Panels:} With e-p and without e-e interaction. 
  {\it Bottom-Left and Bottom-Right Panels:} With e-p and e-e interactions.}
  \label{fig6} 
\end{figure*}

In Figure \ref{fig5}, we present the simulation results for two different combinations of $\gamma_{e,in}^{\prime}$ and $\gamma_{p,in}^{\prime}$ for $\tau_{in} = 4$ when Coulomb (e-p and e-e) interaction is considered and $E_{\gamma,peak}^{\prime} = h\nu_{sa}^{\prime}$ = 0.2 eV, 2 eV and 20 eV (see Equation \ref{Eqn1}). We find that the electron temperature at the end of the simulation is almost unaffected by the choice of $E_{\gamma,peak}^{\prime}$ in the photon seed spectrum. This is because the electrons already cool down to non-relativistic $\gamma_{e}^{\prime}$ as there are enough scatterings with the photons for $\tau_{in} = 4$. The photon spectrum is broader for smaller $E_{\gamma,peak}^{\prime}$ as most of the photons have smaller energy and can thus cool down the electrons more slowly. This results in more photons being upscattered to larger energies as the average number of scatterings per photon is higher. 
As a result, the photon spectrum is shallower above the peak for smaller $E_{\gamma,peak}^{\prime}$: $f_{\nu} \propto \nu^{-0.5}$ for $E_{\gamma,peak}^{\prime}$ = 20 eV, $f_{\nu} \propto \nu^{-0.5}$ for $E_{\gamma,peak}^{\prime}$ = 2 eV and $f_{\nu} \propto \nu^{-0.4}$ for $E_{\gamma,peak}^{\prime}$ = 0.2 eV. 
At photon energies smaller than $E_{\gamma,peak}$, $f_{\nu} \propto \nu^{1.2} - \nu^{1.4}$ which becomes steeper for larger value of $E_{\gamma,peak}^{\prime}$. 
The photons have a lower peak-energy at the end of the simulation as they cool down adiabatically.

In Figure \ref{fig6}, we present the evolution of $\gamma_{e}^{\prime}$ ($\gamma_{p}^{\prime}$) of 3 randomly selected electrons (protons) for $\tau_{in} = 8$, $\gamma_{e,in}^{\prime}$ = 75 and $\gamma_{p,in}^{\prime}$ = 2 when: 1) both e-p and e-e interactions are not considered, 2) only e-p interaction is considered and, 3) both e-p and e-e interactions are considered. The spikes in $\gamma_{e}^{\prime}$ correspond to the instances where the electron interacts either with a proton or a highly energetic photon resulting in a large transfer of energy to the electron. After each such instance, the energy of the electron falls back quickly to non-relativistic values when it upscatters a photon to transfer almost all the kinetic energy that was gained earlier. 
The electrons cool down very fast from $\gamma_{e,in}^{\prime}$ to $\gamma_{e}^{\prime} \sim 1$ as the IC timescale is much smaller than the dynamical timescale $t_{dyn}^{\prime} = R/(\Gamma c)$ and the electron heating timescale $[(\gamma_{e}^{\prime} - 1)m_{e}/(\gamma_{p}^{\prime} - 1)m_{p}]t_{p,Coul}^{\prime}$ for large $\gamma_{e}^{\prime}$ (see Equation \ref{Eqn13}). It can be seen that the 3 electrons experience different number of scatterings which is expected as electrons which are moving towards the photons are more likely to get scattered by the photons than the electrons which are moving away from the photons (Equation \ref{Eqn12}). 

We compare $\gamma_{e}^{\prime}$ at the end of the simulation for the electron which experiences the largest number of scatterings for each of the three cases to find that e-p and e-e interactions do not have very significant effect on $\gamma_{e}^{\prime}$: $\gamma_{e}^{\prime} = 1.048$ without e-p and e-e, $\gamma_{e}^{\prime} = 1.062$ with e-p and without e-e, $\gamma_{e}^{\prime} = 1.059$ with e-p and e-e. The electrons get cooled down faster to small $\gamma_{e}^{\prime}$ ($\sim \gamma_{e}^{\prime}$ without Coulomb) when e-e interaction is included in addition to e-p interaction as $t_{e,Coul}^{\prime}$ is $\sim \beta_{e}/\beta_{p}$ times smaller than $[(\gamma_{e}^{\prime} - 1)m_{e}/(\gamma_{p}^{\prime} - 1)m_{p}]t_{p,Coul}^{\prime}$. The protons have $\gamma_{p}^{\prime} \sim 2$ for $\sim 10^{6}$ scatterings, beyond which their energy drops significantly due to adiabatic cooling. The protons cool down to $\gamma_{p}^{\prime} = 1.123$ for all three cases (irrespective of whether Coulomb interaction is considered) as $t_{ad}^{\prime} \sim t_{dyn}^{\prime}$ is much smaller than $t_{p,Coul}^{\prime}$ for large $R$ towards the end of the simulation ($t_{ad}^{\prime} \propto R$ whereas $t_{p,Coul}^{\prime} \propto R^{2}$).

\begin{figure*}
\gridline{\fig{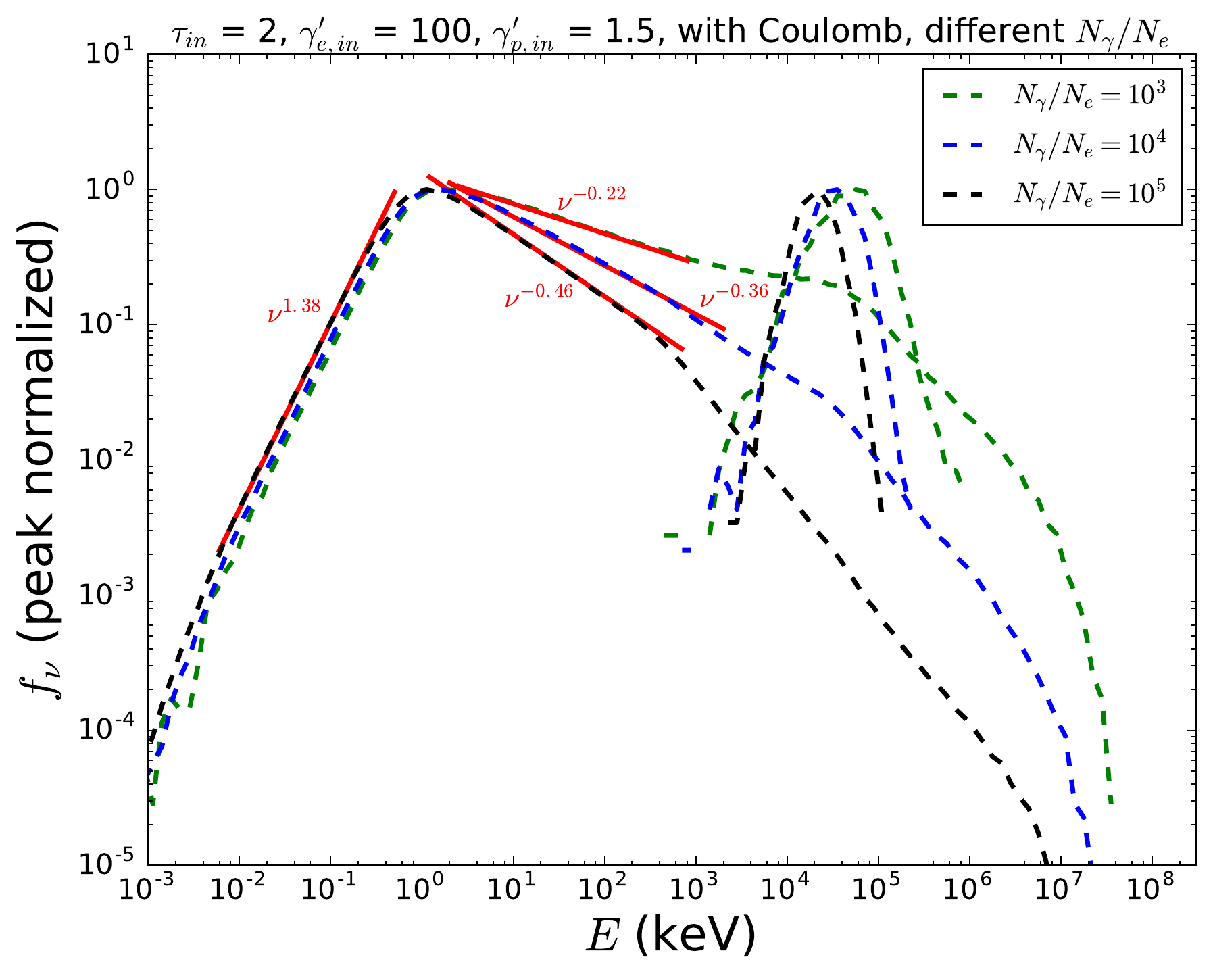}{0.5\textwidth}{}
          \fig{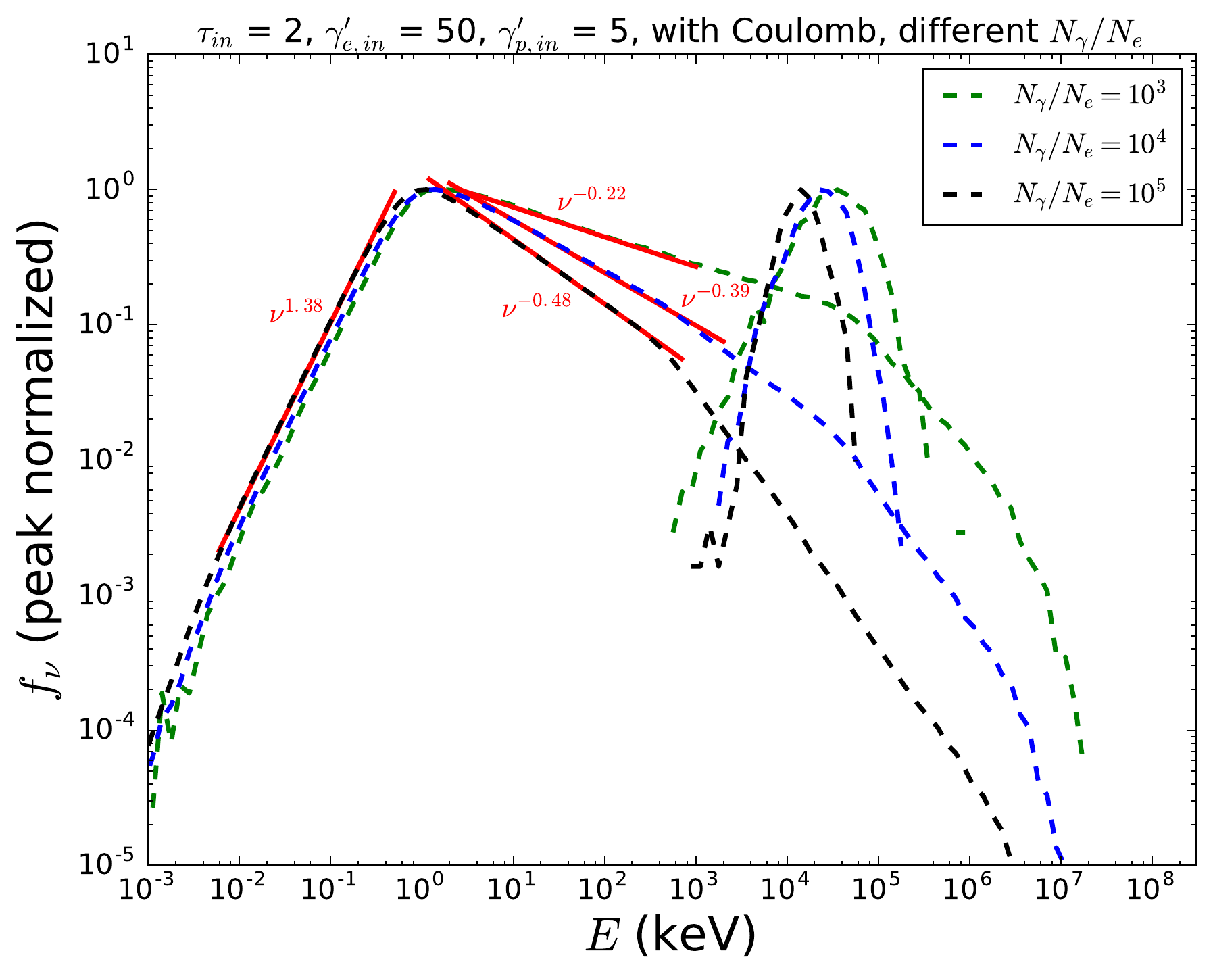}{0.5\textwidth}{}
          }  \vspace{-2em}
  \caption{Comparison of simulation results for photons with seed spectrum given by Equation \ref{Eqn1}, $\tau_{in}$ = 2 and with Coulomb interaction (e-p and e-e) for different $N_{\gamma}/N_{e} = 3\times10^{6}/3\times10^{3}, 3\times10^{7}/3\times10^{3}\: \rm{and}\: 3\times10^{8}/3\times10^{3}$.
 {\it Left Panel:} For $\gamma_{p,in}^{\prime} = 1.5$ and $\gamma_{e,in}^{\prime}$ = 100. 
 {\it Right Panel:} For $\gamma_{p,in}^{\prime} = 5$ and $\gamma_{e,in}^{\prime}$ = 50.}
  \label{fig7} 
\end{figure*}

\begin{figure*}
\gridline{\fig{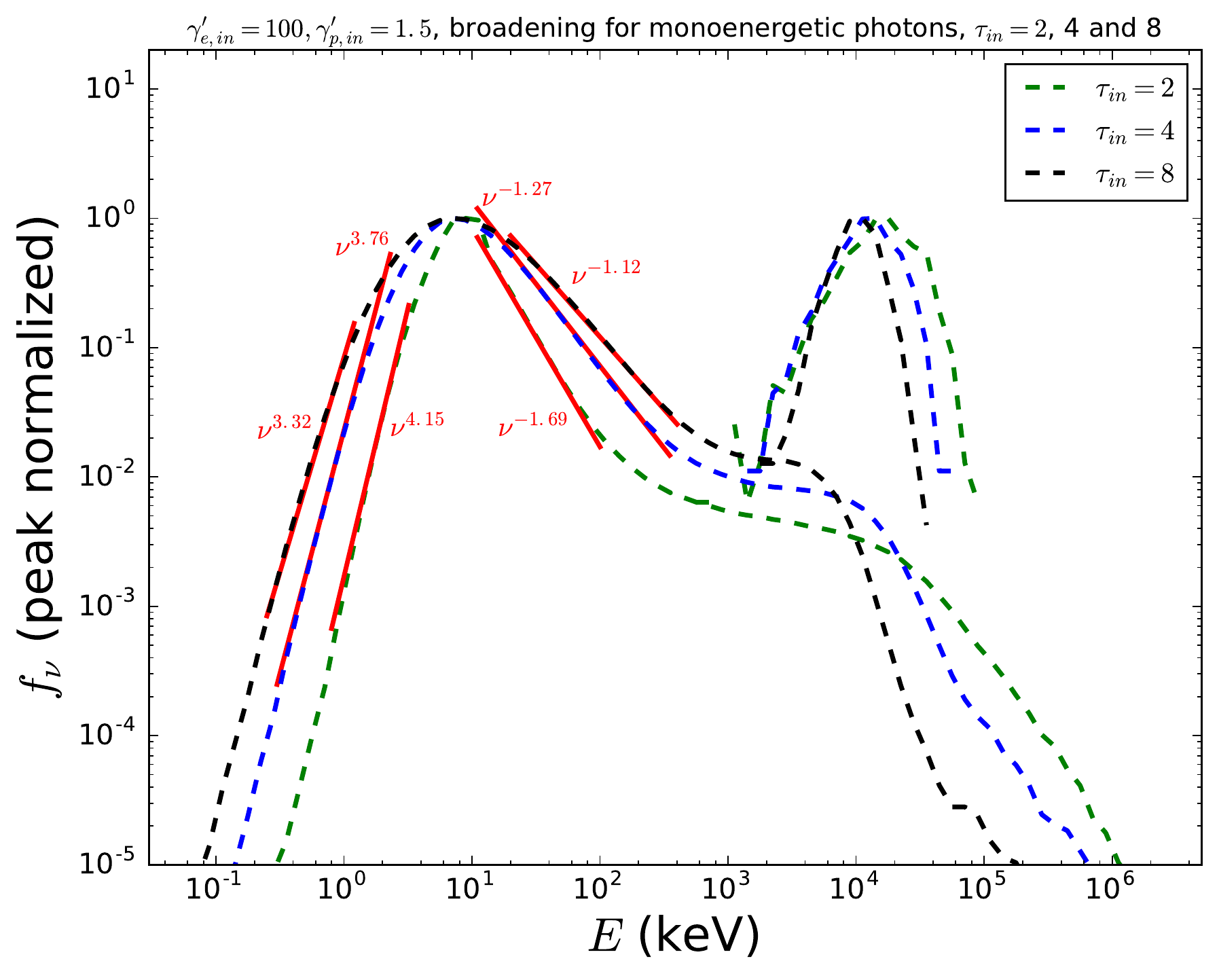}{0.5\textwidth}{}
          \fig{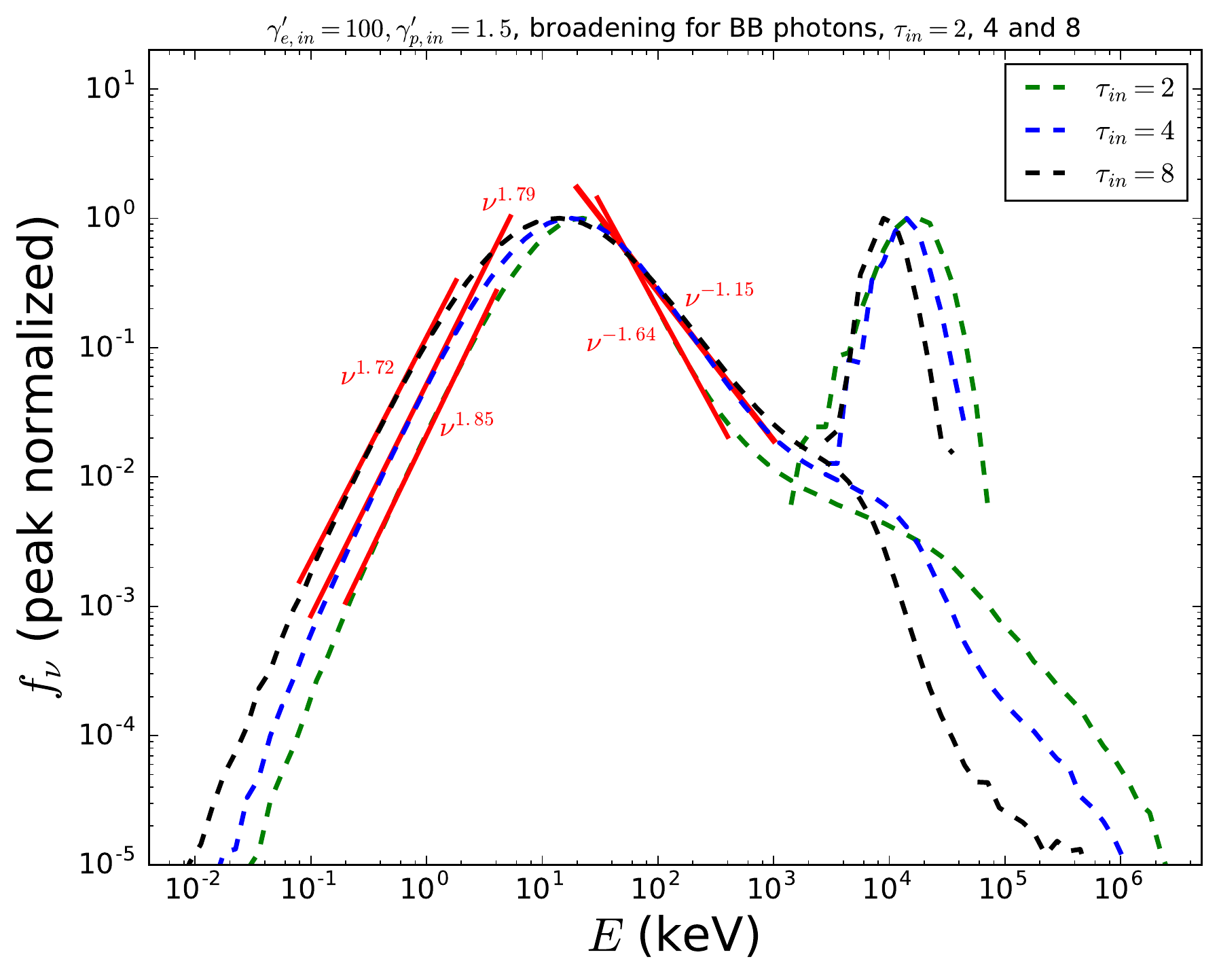}{0.5\textwidth}{}
          }  \vspace{-2em}   
  \caption{Broadening of seed photon spectrum for large $\tau_{in} = 2$, 4 and 8, with Coulomb interaction (e-p and e-e), $\gamma_{e,in}^{\prime} = 100$, $\gamma_{p,in}^{\prime} = 1.5$, $N_{\gamma} = 10^{8}$ and $N_{e} = 10^{3}$.
{\it Left Panel:} For monoenergetic photon seed spectrum with $E_{\gamma,peak}^{\prime} = 20$ eV.
{\it Right Panel:} For blackbody (BB) photon seed spectrum with $E_{\gamma,peak}^{\prime} = 20$ eV.}
\label{fig8} 
\end{figure*}

In Figure \ref{fig7}, we present the simulation results for two different combinations of $\gamma_{e,in}^{\prime}$ and $\gamma_{p,in}^{\prime}$ for $\tau_{in} = 2$ when Coulomb (e-p and e-e) interaction is considered and $N_{\gamma}/N_{e} = 3\times10^{6}/3\times10^{3}$, $3\times10^{7}/3\times10^{3}$ and $3\times10^{8}/3\times10^{3}$. We find that the electrons are considerably hotter at the end of the simulation for smaller $N_{\gamma}/N_{e}$: $\gamma_{e}^{\prime} = 1.391\: (1.261)$ for $N_{\gamma}/N_{e} = 10^3$, $\gamma_{e}^{\prime} = 1.196\: (1.144)$ for $N_{\gamma}/N_{e} = 10^4$ and $\gamma_{e}^{\prime} = 1.130\: (1.091)$ for $N_{\gamma}/N_{e} = 10^5$ when $\gamma_{e,in}^{\prime} = 100$ and $\gamma_{p,in}^{\prime} = 1.5$ ($\gamma_{e,in}^{\prime} = 50$ and $\gamma_{p,in}^{\prime} = 5$).
This is expected as the electrons cool down faster when there are more photons available to get upscattered. As a result, the photon spectrum becomes shallower above $E_{\gamma,peak}$ as more photons get upscattered by the slowly cooling electrons to higher energies for smaller $N_{\gamma}/N_{e}$: $f_{\nu} \propto \nu^{-0.5}$ for $N_{\gamma}/N_{e} = 10^{5}$, $f_{\nu} \propto \nu^{-0.4}$ for $N_{\gamma}/N_{e} = 10^{4}$ and $f_{\nu} \propto \nu^{-0.2}$ for $N_{\gamma}/N_{e} = 10^{3}$ from the peak energy upto $\sim 10^{3}$ keV. 

In Figure \ref{fig8}, we present the simulation results for the broadening of mono-energetic and blackbody seed photon spectra with $E_{\gamma,peak}^{\prime} = 20$ eV, $\gamma_{e,in}^{\prime} = 100$ and $\gamma_{p,in}^{\prime} = 1.5$ when $\tau_{in} = 2$, 4 and 8. As expected, the electrons have smaller energy at the end of the simulation for larger $\tau_{in}$: $\gamma_{e,in}^{\prime} = 1.130$ ($\gamma_{e,in}^{\prime} = 1.117$) for $\tau_{in} = 2$, $\gamma_{e,in}^{\prime} = 1.065$ ($\gamma_{e,in}^{\prime} = 1.091$) for $\tau_{in} = 4$ and $\gamma_{e,in}^{\prime} = 1.052$ ($\gamma_{e,in}^{\prime} = 1.052$) for $\tau_{in} = 8$ for mono-energetic (blackbody) seed photons. The photon spectrum becomes considerably broader with increasing $\tau_{in}$: $f_{\nu} \propto \nu^{4.2}$ ($f_{\nu} \propto \nu^{1.9}$) for $\tau_{in} = 2$ to $f_{\nu} \propto \nu^{3.3}$ ($f_{\nu} \propto \nu^{1.7}$) for $\tau_{in} = 8$ below $E_{\gamma,peak}$ and $f_{\nu} \propto \nu^{-1.7}$ ($f_{\nu} \propto \nu^{-1.6}$) for $\tau_{in} = 2$ to $f_{\nu} \propto \nu^{-1.1}$ ($f_{\nu} \propto \nu^{-1.2}$) for $\tau_{in} = 8$ above $E_{\gamma,peak}$ for mono-energetic (blackbody) seed photons. Unlike previous simulations, the power-law above $E_{\gamma,peak}$ only extends up to $E_{\gamma} \sim 10^{2}$ keV. 

To summarize, we studied the effect of $\gamma_{e,in}^{\prime}$, $\gamma_{p,in}^{\prime}$, $\tau_{in}$, $E_{\gamma,peak}^{\prime}$, $N_{\gamma}/N_{e}$
and Coulomb (e-p and e-e) interaction on the output spectrum of the photons and the electrons. In Figure \ref{fig2}, we show that $\gamma_{e,in}^{\prime}$ and $\gamma_{p,in}^{\prime}$ do not have any significant effect on $E_{\gamma,peak}^{\prime}$ and the power-law above peak energy except that the high-energy tails in the photon spectrum extend to larger energies for larger $\gamma_{e,in}^{\prime}$. 
In Figure \ref{fig4}, we find that increasing $\tau_{in}$ slightly flattens the photon spectrum at low energies although $f_{\nu}$ drops faster at higher energies. In Figure \ref{fig5}, we find that the peak energy of the seed photon spectrum determines the peak energy and shape of the output photon spectrum. In Figure \ref{fig6}, we track $\gamma_{e}^{\prime}$ to establish that e-p and e-e interactions do not affect the electron energies significantly which is in good agreement with the previous simulations (Figures \ref{fig2} - \ref{fig5}). In Figure \ref{fig7}, we find that although $E_{\gamma,peak}^{\prime}$ is unaffected by the decrease in $N_{\gamma}/N_{e}$, $f_{\nu}$ increases significantly above $E_{\gamma,peak}^{\prime}$ resulting in shallower photon spectrum. In Figure \ref{fig8}, we find that the output photon spectrum is broadened for large $\tau_{in}$ irrespective of the choice of the seed photon spectrum. This implies that the output photon spectrum for high $\tau_{in} \sim$ few tens - hundred will be in good agreement with the observed Band spectrum. 


\section{Discussion of results}
\label{Discussions}
In this section, we first discuss the simulation parameters that significantly affect the output photon spectrum. Then we discuss the energy constraint that the electrons must satisfy in order to transfer enough energy to the photons so that a power-law can be produced above $E_{\gamma,peak}$. Although this constraint is a necessary condition, it is not a sufficient condition to ensure a power-law spectrum for photons at high energies (\citealt{Santana16}). Next, we discuss the evolution of energies for the photons, protons and the electrons due to processes such as Comptonization, adiabatic cooling and Coulomb interaction (e-p and e-e) during the expansion of the relativistic jet. We also evaluate the equilibrium $\gamma_{e}^{\prime}$ when the electrons are cooled due to IC and are heated by the protons due to e-p interaction. Lastly, we discuss the effect of $N_{\gamma}/N_{e}$ on our simulation results.

\subsection{Effect of simulation parameters on the output spectrum}
In our simulations, the parameters that mainly affect the output photon spectrum are $\gamma_{e,in}^{\prime}$, $\gamma_{p,in}^{\prime}$, $E_{\gamma, peak}^{\prime} = h\nu_{sa}^{\prime}$, $\tau_{in}$ and $N_{\gamma}/N_{e}$. 
The e-p interaction slightly elevates the energy of the electrons but it does not change the photon spectrum and the proton energies appreciably. The e-e interaction plays an important role in redistributing energy among the electrons after they are heated by the e-p interactions.


$\gamma_{e,in}^{\prime}$ and $\gamma_{p,in}^{\prime}$ determine the amount of energy that the electrons can transfer to the photons through Comptonization and the amount of energy that the protons can transfer to the electrons through e-p interactions, respectively. Higher $\gamma_{p,in}^{\prime}$ can also result in more energetic photons as the electrons will gain more energy from the protons to transfer it to the photons. However, for most part of the simulations, $[(\gamma_{e}^{\prime} - 1)m_{e}/(\gamma_{p}^{\prime} - 1)m_{p}]t_{p,Coul}^{\prime}$ is of the same order as IC timescale in the jet-comoving frame which is given by,
\begin{equation}
\label{Eqn13}
t_{IC}^{\prime} = \frac{3}{4} \frac{(\gamma_{e}^{\prime} - 1)m_{e}c}{U_{\gamma}^{\prime}\sigma_{T}\gamma_{e}^{\prime 2} \beta_{e}^{\prime 2}}
\end{equation}
where $U_{\gamma}^{\prime} = L_{\gamma}/(4\pi R^{2} \Gamma^{2} c)$ is the radiation energy density. Hence, the
electrons attain an equilibrium $\gamma_{e}^{\prime}$ after a certain number of scatterings and Coulomb interaction is relatively unimportant in determining the shape of the output photon spectrum. 

$E_{\gamma, peak}^{\prime}$ is also an important parameter that affects the shape of the output photon spectrum. However for almost all our simulations (except Figure \ref{fig5}), we fix the seed photon spectrum (as given by Equation \ref{Eqn1}) to study the effect of other parameters and interactions better. As a photon can be upscattered to an energy $\sim E_{\gamma, peak}^{\prime} \Gamma \gamma_{e,in}^{\prime \:2}$ after one scattering, more energetic photons (with higher $E_{\gamma, peak}^{\prime}$) cool the electrons faster after multiple scatterings. 
We do not consider electron-positron pair production for our simulations as $E_{\gamma, peak}^{\prime} \sim 0.2 - 20$ eV is much less than the rest mass energy of the electrons in the jet-comoving frame (see Appendix \ref{AppendixC}, for more details).

The average number of scatterings experienced by a photon before escaping the photosphere is $\sim 2\tau_{in}$ (\citealt{Begue13}) and hence $\tau_{in}$ also determines the shape of the output photon spectrum. 
For larger $\tau_{in}$, the electrons and the protons cool down more adiabatically (see Equations \ref{Eqn4} and \ref{Eqn5}). The photons get scattered multiple times thus increasing the probability of different photons getting scattered different number of times before escaping the photosphere. This results in broadening of the photon spectrum and the output photon spectrum looks shallower below $E_{\gamma,peak}$. This broadening of the output photon spectrum at large $\tau_{in}$ is independent of the choice of the seed photon spectrum (see Figure \ref{fig8}).

Another parameter which affects the photon and electron energies is $N_{\gamma}/N_{e}$. For smaller $N_{\gamma}/N_{e}$, there are more electrons to upscatter the photons to higher energies. Moreover, $N_{e} = N_{p}$ implies that there are more protons to transfer energy to the electrons. Hence, the output photon spectrum has more photons at higher energies resulting in a shallower spectrum above $E_{\gamma,peak}$.

It should be noted that unlike previous simulations (\citealt{LB10,Santana16}), we do not re-accelerate the electrons back to their initial distribution after every few scattering events. Rather the electrons are redistributed to MB distribution whose peak temperature is determined using $\gamma_{e,avg}^{\prime}$ after each scattering event. We do not consider any external dissipation events for electron heating except the energy transfer from the protons to the electrons.

\subsection{Energy constraint for power-law above $E_{\gamma,peak}^{\prime}$}
Now we discuss the constraint that $\gamma_{e,in}^{\prime}$ and $\gamma_{p,in}^{\prime}$ need to satisfy in order to have a power-law spectrum above $E_{\gamma,peak}^{\prime}$. The total initial kinetic energy of the electrons and the protons at the beginning of the simulation is $(\gamma_{e,in}^{\prime} - 1) N_{e}m_e c^2$ and $(\gamma_{p,in}^{\prime} - 1) N_{p}m_p c^2$, respectively. The energy available to the electrons should atleast be as large as the energy gain that is required by the photons to populate the high-energy tail. The energy transferred from the protons to the electrons in the course of the jet expansion is $\sim (t_{dyn}^{\prime}/t_{p,Coul}^{\prime}) (\gamma_{p,in}^{\prime} - 1) N_{p}m_p c^2$. In order to have a power-law spectrum above $E_{\gamma,peak}^{\prime}$, the energy of a fraction $\sim f$ of the photons near peak-energy $E_{\gamma,peak}^{\prime}$ has to increase by a factor $\sim f$. Assuming that most of the photons have energies close to the photon peak-energy $E_{\gamma,peak}^{\prime}$, the electron $\gamma_{e,in}^{\prime}$ and the proton $\gamma_{p,in}^{\prime}$ should satisfy the energy constraint given by
\begin{equation}
\label{Eqn14}
(\gamma_{e,in}^{\prime} - 1) N_{e}m_e c^2 + \frac{t_{dyn}^{\prime}}{t_{p,Coul}^{\prime}} (\gamma_{p,in}^{\prime} - 1) N_{p}m_p c^2\gtrsim f \frac{N_{\gamma}}{f} E_{\gamma,peak}^{\prime}
\end{equation}
For our simulations, $E_{\gamma,peak}^{\prime}$ = 2 eV, $N_{\gamma} = 10^8$ and $N_{p} = N_{e} = 10^3$. $t_{dyn}^{\prime}$ and $t_{p,Coul}^{\prime}$ are evaluated when most of the scatterings occur with $\gamma_{e}^{\prime} \sim 1$ and $\gamma_{p}^{\prime} \sim \gamma_{p,in}^{\prime}$ (see Figure \ref{fig6}). For this choice of parameters, the above condition is satisfied for $\gamma_{e,in}^{\prime} \sim 25 - 100$ and $\gamma_{p,in}^{\prime} \sim 1.5 - 10$ that we have considered. This explains the power-law spectrum from $E_{\gamma,peak}$ upto $\sim 10^3$ keV in all our simulations.

\subsection{Energy evolution for the photons, protons and electrons}
Now we discuss the evolution of energy for the photons, protons and electrons to explain our simulation results.

\subsubsection{Photons}
The photons gain energy from the electrons through Comptonization and cool due to adiabatic expansion. The IC timescale is much smaller compared to $t_{ad}^{\prime} \sim t_{dyn}^{\prime}$ until the electrons cool down to non-relativistic energies ($\gamma_{e}^{\prime} \sim 1)$. Although some photons are upscattered to high energies through Comptonization when the electrons are hot, the peak of the photon spectrum is affected only by adiabatic cooling and not IC cooling. This is expected as most of the scatterings occur after the electrons already cool down to non-relativistic energies making Comptonization unimportant for determining $E_{\gamma,peak}^{\prime}$ in the output photon spectrum.

The peak-energy of the output photon spectrum can be obtained using Equations \ref{Eqn3} and \ref{Eqn6},
\begin{equation}
\frac{E_{\gamma,peak,f}^{\prime}}{E_{\gamma,peak,i}^{\prime}} = \left(\frac{R_{ph}}{R_{in}}\right)^{-2/3} = \tau_{in}^{-2/3}
\end{equation}
where $E_{\gamma,peak,i}^{\prime}$ ($E_{\gamma,peak,f}^{\prime}$) is the peak-energy of the initial (final) photon spectrum in the jet-comoving frame. This is in good agreement with our simulation results for different $\tau_{in}$ and $E_{\gamma,peak}^{\prime}$ in Figures \ref{fig4} and \ref{fig5}. 

\subsubsection{Protons}
The protons lose energy to the electrons through e-p interaction in addition to cooling adiabatically as the jet expands. The proton cooling timescale $t_{p,Coul}^{\prime}$ is much larger compared to $t_{ad}^{\prime}$ when the electrons are relativistic (see Equation \ref{Eqn7}). After the electrons cool down to non-relativistic energies, $t_{p,Coul}^{\prime} \propto (\gamma_{p}^{\prime} - 1)(\beta_{p}^{\prime}/n_{e}^{\prime}) \propto R^2$ which increases faster compared to $t_{ad}^{\prime} \propto R$ with the expansion of the jet. Thus, e-p interaction is relatively unimportant in determining the final energy of the protons which is actually determined by adiabatic cooling. From Figure \ref{fig6}, we can see that $\gamma_{p}^{\prime} \sim 2$ for most part of the simulation. Thus, we can write using Equation \ref{Eqn5},
\begin{equation}
\frac{\gamma_{p,f}^{\prime} - 1}{\gamma_{p,i}^{\prime} - 1} \sim \left(\frac{R_{ph}}{R_{in}}\right)^{-1} = \tau_{in}^{-1}
\end{equation}
as $\gamma_{ad,p}^{\prime} \sim 3/2$. For $\tau_{in} = 8$ and $\gamma_{p,in}^{\prime} = 2$ as used in the simulations in Figure \ref{fig6}, $\gamma_{p,f}^{\prime} = 1.125$ which is in very good agreement with the $\gamma_{p}^{\prime}$ value that we find by tracking the protons in Figure \ref{fig6}.

\subsubsection{Electrons}
While the electrons gain energy from the protons through e-p interaction, they also lose energy due to adiabatic expansion of the jet and Comptonization. Comptonization of electrons no longer decreases the energy of the electrons significantly after $\gamma_{e}^{\prime}$ drops to $\gamma_{e,Comp}^{\prime} = 1 + 1/(8\tau_{in})$ (\citealt{Santana16}). 
Here we estimate the change in energy of the electrons due to the three processes: adiabatic cooling, Comptonization and e-p interaction, after the electrons have cooled down to $\gamma_{e,Comp}^{\prime}$ to explain our simulation results in Figure \ref{fig6}. 

The evolution of $\gamma_{e}^{\prime}$ due to adiabatic cooling is given by Equation \ref{Eqn4}. After the electrons have already cooled down to $\gamma_{e}^{\prime} \sim 1$ the energy change due to adiabatic cooling is,
\begin{equation}
\frac{\gamma_{e,f}^{\prime} - 1}{\gamma_{e,i}^{\prime} - 1} \sim \left(\frac{R_{ph}}{R_{in}}\right)^{-4/3} = \tau_{in}^{-4/3}
\end{equation}
as $\gamma_{ad,e}^{\prime} \sim 5/3$. For $\tau_{in} = 8$ and $\gamma_{e,i}^{\prime} = \gamma_{e,Comp}^{\prime} \sim 1.016$ as used in simulations in Figure \ref{fig6}, $\gamma_{e,f}^{\prime} \sim  1.001$. To estimate the change in $\gamma_{e}^{\prime}$ due to Comptonization and Coulomb heating by the protons, we first evaluate the corresponding timescales along with the dynamical timescale (all timescales averaged over $R$) for $\tau_{in} = 8$. For $\gamma_{e}^{\prime} = \gamma_{e,Comp}^{\prime}$ and $\gamma_{p}^{\prime} \sim 2$, the $R$-averaged timescales are,
\begin{equation}
\langle t_{dyn}^{\prime}\rangle_{R} = \left\langle\frac{R}{\Gamma c}\right\rangle_{R} = 0.05\:\rm{s}
\end{equation}
\begin{equation}
\langle t_{IC}^{\prime}\rangle_{R} = \left\langle\frac{3}{8}\frac{m_{e}c}{U_{\gamma}^{\prime}\sigma_{T}\gamma_{e,Comp}^{\prime}}\right\rangle_{R} = 1.25\times10^{-4}\:\rm{s}
\end{equation}
\begin{equation}
\langle t_{Coul}^{\prime}\rangle_{R} = \left\langle\frac{(\gamma_{e,Comp}^{\prime} - 1)m_{e}c^{2}}{5\times10^{-19}n_{e}^{\prime}}\beta_{p}^{\prime}\right\rangle_{R} = 2.82\times10^{-5}\:\rm{s}
\end{equation}
The final energy of the electrons after $t= \langle t_{dyn}^{\prime}\rangle_{R}$ due to IC is given by,
\begin{equation}
E_{e,f,IC}^{\prime} = E_{e,i}^{\prime} e^{-\langle t_{dyn}^{\prime}\rangle_{R}/\langle t_{IC}^{\prime}\rangle_{R}}
\end{equation}
which reduces to,
\begin{equation}
\gamma_{e,f,IC}^{\prime} = 1 + (\gamma_{e,Comp}^{\prime} - 1) e^{-400} \sim 1
\end{equation}
$\gamma_{e}^{\prime}$ of the electrons after $t = \langle t_{dyn}^{\prime}\rangle_{R}$ due to Coulomb heating by the protons is,
\begin{equation}
\gamma_{e,f,Coul}^{\prime} = 1 + (\gamma_{e,Comp}^{\prime} - 1) e^{\langle t_{dyn}^{\prime}\rangle_{R}/\langle t_{Coul}^{\prime}\rangle_{R}} \gg \gamma_{e,Comp}^{\prime}
\end{equation}
Thus, we find that the electron heating due to the Coulomb interaction is faster than the rate of Comptonization of the electrons at smaller values of $\tau \sim 1$ i.e. towards the end of the simulation. This explains our results in Figure \ref{fig6} as $\gamma_{e}^{\prime}$ obtained at the end of the simulation is higher when e-p interaction is considered.


\subsection{Equilibrium $\gamma_{e}^{\prime}$ of electrons}
In this subsection, we evaluate $\gamma_{e}^{\prime}$ after the electrons reach equilibrium due to Coulomb heating and Comptonization. As the timescale for adiabatic cooling of electrons $\sim t_{dyn}^{\prime}$ is much longer as compared to $t_{IC}^{\prime}$ and $t_{Coul}^{\prime}$, we can neglect adiabatic cooling while considering the equilibrium of the electrons. Equating the IC energy loss rate with the Coulomb energy gain rate for the electrons gives, 
\begin{equation}
\frac{5\times10^{-19}n_{e}^{\prime}\beta_{p}^{\prime 2}}{8.3\times10^{-15}[(\gamma_{e}^{\prime}-1)m_{e}c^{2}/k_{B}]^{3/2} + \beta_{p}^{\prime 3}} = \frac{4}{3}U_{\gamma}^{\prime}\sigma_{T}(\gamma_{e}^{\prime 2} - 1)c
\end{equation}
Using the expressions for $n_{e}^{\prime}$, $U_{\gamma}^{\prime}$ and $\gamma_{p}^{\prime} = 1.123$, we obtain
\begin{equation}
39.07 (\gamma_{e}^{\prime} - 1)^{3/2} + 1 = \frac{0.273}{\gamma_{e}^{\prime 2} - 1}
\end{equation} 
which gives $\gamma_{e}^{\prime} = 1.074$. Thus, the equilibrium $\gamma_{e}^{\prime}$ is close to $\gamma_{e}^{\prime} = 1.062$ obtained in Figure \ref{fig6} when e-p interaction is considered. The equilibrium $\gamma_{e}^{\prime}$ is slightly higher than $\gamma_{e}^{\prime}$ obtained at the end of the simulations for Figure \ref{fig6} which is expected as we neglect adiabatic cooling and e-e interaction for our equilibrium calculations.

In our analysis, we have assumed that the electrons always cool due to IC and neglect the possibility that an energetic photon can transfer energy back to the electrons. However, there are about $\sim 10$ instances in each of the three cases (without e-p and e-e, with e-p and with e-p and e-e) when the electron energy increases to $\gamma_{e}^{\prime} \sim 2$. As a result, more photons are upscattered to higher energies and the power-law $f_{\nu} \propto \nu^{-0.5}$ extends to $\sim 10^3$ keV for almost all our simulations. In addition, the Compton-Y parameter for sub-relativistic electrons (\citealt{RL79}) is,
\begin{equation}
Y = 2\tau_{in}\times \frac{4k_{B}T_{e}^{\prime}}{m_{e}c^{2}} \sim 8\tau_{in}\times (\gamma_{e}^{\prime} - 1)
\end{equation}
which is $\sim 4$ at the end of the simulation for $\gamma_{e}^{\prime} \sim 1.062$ and $\tau_{in} = 8$ (see Figure \ref{fig6}) - large enough to upscatter most of the photons by a factor of 2 in energy and populate the high-energy tail of the photon spectrum. The considerably large value of Compton-Y parameter accounts for the upscattering of photons near $E_{\gamma,peak}$ to the high energy power-law region of the photon spectrum.

\subsection{Effect of $N_{\gamma}/N_{e}$ on simulation results}
In this subsection, we discuss the simulation results in Figure \ref{fig7} which were performed at $\tau_{in} = 2$ and different values of $N_{\gamma}/N_{e} = 10^{3}, 10^{4}\: \rm{and\: 10^{5}}$ for two different combinations of $\gamma_{e,in}^{\prime}$ and $\gamma_{p,in}^{\prime}$. The number of electrons $N_{e} = 3\times10^3$ is kept constant for the simulations and $N_{\gamma}\: (=3\times10^6, 3\times10^7 \:\rm{and \: 3\times10^8})$ is varied. We find that the electrons are hotter and the photon spectrum is shallower for smaller $N_{\gamma}/N_{e}$ for both combinations of $\gamma_{e,in}^{\prime}$ and $\gamma_{p,in}^{\prime}$ considered in Figure \ref{fig7}. 
Rewriting Equation \ref{Eqn14} using Equation \ref{Eqn7} and the fact that $N_{p} = N_{e}$,
\begin{equation}
(\gamma_{e,in}^{\prime} - 1) m_{e} c^2 + \frac{5\times10^{-19} n_{e}^{\prime}}{\beta_{p,in}^{\prime}} \frac{R}{\Gamma c} \gtrsim \frac{N_{\gamma}}{N_{e}} E_{\gamma,peak}^{\prime}
\end{equation}
Thus, for a given $\gamma_{e,in}^{\prime}$, $\gamma_{p,in}^{\prime}$ and $E_{\gamma,peak}^{\prime}$, the electrons cannot transfer enough energy to the photons to populate the higher energy power-law tail for larger $N_{\gamma}/N_{e}$. As a result, the photon spectrum falls down faster at higher energies for larger values of $N_{\gamma}/N_{e}$. Our simulations show that the photon spectrum is significantly affected by the choice of $N_{\gamma}/N_{e}$ and it is important to perform the simulations with realistic values of $N_{\gamma}/N_{e} = 10^5$. For all three values of $N_{\gamma}/N_{e}$ we have more photons just above the peak photon energy as compared to previous simulations (\citealt{LB10, Santana16}), which is due to
smaller $E_{\gamma,peak}^{\prime}$ resulting in slower cooling of the electrons by the photons.

\section{Conclusions}
\label{Conclusions}
We studied photospheric emission for GRB prompt emission using a MC code with photon to electron number ratio $N_{\gamma}/N_{e} = 10^5$, which is close to the expected value for a typical GRB if the radiation efficiency is $\sim 10\%$.
Our objective was to find out whether photospheric emission can explain the observed non-thermal low ($f_{\nu} \propto \nu^{0}$) and high-energy ($f_{\nu} \propto \nu^{-1.2}$) spectrum of GRB prompt emission. For all our simulations, we considered Comptonization of seed photons with synchrotron spectrum in fast cooling regime. The electrons are continuously heated by the mono-energetic protons as the electrons interact with the protons (e-p) and other electrons (e-e) through Coulomb interaction. In all our simulations, we also consider the energy change for electrons, photons and protons due to the adiabatic expansion of the jet. We find that the output photon spectrum exhibits a power-law
extending upto $\sim 10^3$ keV from $E_{\gamma,peak}^{\prime}$ for the parametric space of initial electron energy ($\gamma_{e,in}^{\prime}$), initial proton energy ($\gamma_{p,in}^{\prime}$) and initial optical depth ($\tau_{in}$) that we consider in this work. 

We find that the output photon spectrum becomes slightly shallower below $E_{\gamma,peak}^{\prime}$ as the initial optical depth $\tau_{in}$ increases.  
This is expected as photons get scattered by electrons different number of times before escaping out of the photosphere as $\tau_{in}$ increases. This can possibly result in an output photon spectrum which is in good agreement with the observed low-energy spectrum $f_{\nu} \propto \nu^{0}$ of the prompt emission, especially at large $\tau_{in} \sim$ few tens-hundred.
The flattening of the output photon spectrum below $E_{\gamma,peak}^{\prime}$ for large $\tau_{in}$ is independent of the choice of seed photon spectrum.
We find that the peak energy and shape of the output photon spectrum is also determined by the peak energy $E_{\gamma,peak}^{\prime}$ of the seed photon spectrum. The peak energy in the output spectrum reduces by a factor $\sim \tau_{in}^{-2/3}$ compared to the seed spectrum because of adiabatic cooling of photons. As expected, the photon spectrum is broader around the peak energy for smaller $E_{\gamma,peak}^{\prime}$ because the photons are less energetic in the jet-comoving frame and can cool the electrons more slowly resulting in more scatterings. 

We track the electrons and the protons to study the effect of Coulomb (e-p and e-e) interaction on the electron and proton energies and the output photon spectrum. We find that the electron energies are slightly elevated in the presence of Coulomb interaction and the protons cool down considerably by the end of the simulation due to adiabatic expansion of the jet for the optical depths that we consider. The presence of Coulomb interaction does not affect $E_{\gamma,peak}^{\prime}$ and the shape of the output photon spectrum (both below and above $E_{\gamma,peak}$) in general. We evaluate $\gamma_{e}^{\prime}$ at equilibrium due to IC and e-p interactions and find that Compton-Y parameter $\sim 4$ at the end of the simulation - which is large enough to populate the high-energy power-law tail of the photon spectrum. 

We also performed simulations for different $N_{\gamma}/N_{e}$ and found that the photon spectrum becomes shallower above $E_{\gamma,peak}^{\prime}$ and does not exhibit power-law tail at high energies for smaller $N_{\gamma}/N_{e}$. This shows the importance of performing simulations with realistic $N_{\gamma}/N_{e}$ and thus radiation efficiency $\eta$. We find that the Comptonization of seed photons with synchrotron spectrum in fast cooling regime cannot explain the high energy power-law dependence ($f_{\nu} \propto \nu^{-1.2}$) and the peak energy of the observed GRB prompt emission spectrum. However, $f_{\nu} \propto \nu^{0}$ for the photon spectrum below $E_{\gamma,peak}$ can be successfully explained using fast cooling synchrotron seed photon spectrum at very large optical depths ($\tau_{in} \sim 100$). 


\section*{Acknowledgments}
MB would like to thank Milos Milosavljevic for providing the computational facilities required for this work.


\begin{appendix}

\section{Initialization of photon energy}
\label{AppendixA}
Here we describe the algorithm that we implemented to draw seed photons from synchrotron spectrum for fast cooling electrons. The energy distribution is given by Equation \ref{Eqn1} with break energies $E_{\gamma,1}^{\prime} = h\nu_{l}$, $E_{\gamma,2}^{\prime} = h\nu_{ac}$, $E_{\gamma,3}^{\prime} = h\nu_{sa}$, $E_{\gamma,4}^{\prime} = h\nu_{m}$ and $E_{\gamma,5}^{\prime} = h\nu_{u}$. We denote the spectral indices between the break energies using $p_{1} = -1.0$, $p_{2} = 2.0/3.0$, $p_{3} = 1.5$ and $p_{4} = 4.5/2.0$ where the photon spectrum is given by $f_{\nu} \propto \nu^{1-p}$. 

We first evaluate, 
\begin{equation}
C_{1} = \frac{E_{1}^{1-p_{1}} - E_{2}^{1-p_{1}}}{p_{1} - 1} \nonumber
\end{equation}
\begin{equation}
C_{2} = \left(\frac{E_{2}^{1-p_{2}} - E_{3}^{1-p_{2}}}{p_{2} - 1}\right) E_{2}^{p_{2} - p_{1}} \nonumber 
\end{equation}
\begin{equation}
C_{3} = \left(\frac{E_{3}^{1-p_{3}} - E_{4}^{1-p_{3}}}{p_{3} - 1}\right) E_{2}^{p_{2} - p_{1}}E_{3}^{p_{3} - p_{2}} \nonumber
\end{equation}
\begin{equation}
C_{4} = \left(\frac{E_{4}^{1-p_{4}} - E_{5}^{1-p_{4}}}{p_{4} - 1}\right) E_{2}^{p_{2} - p_{1}}E_{3}^{p_{3} - p_{2}}E_{4}^{p_{4} - p_{3}}  \nonumber
\end{equation}
to find $K_{1} = C_{1}/(C_{1}+C_{2}+C_{3}+C_{4})$, $K_{2} = (C_{1}+C_{2})/(C_{1}+C_{2}+C_{3}+C_{4})$ and $K_{3} = (C_{1}+C_{2}+C_{3})/(C_{1}+C_{2}+C_{3}+C_{4})$. Next, we draw two random numbers $\xi_{1}$ and $\xi_{2}$ between 0 and 1 to set
\begin{equation}
E_{\gamma}^{\prime} = \left\{
\begin{array}{ll}
[\xi_{2}(E_{\gamma,2}^{\prime 1-p_{1}} - E_{\gamma,1}^{\prime 1-p_{1}}) + E_{\gamma,1}^{\prime 1-p_{1}}]^{1/(1-p_{1})}, & 0 < \xi_{1} < K_{1}\\[3pt] \relax
[\xi_{2}(E_{\gamma,3}^{\prime 1-p_{2}} - E_{\gamma,2}^{\prime 1-p_{2}}) + E_{\gamma,2}^{\prime 1-p_{2}}]^{1/(1-p_{2})}, & K_{1} < \xi_{1} < K_{2}\\[3pt] \relax
[\xi_{2}(E_{\gamma,4}^{\prime 1-p_{3}} - E_{\gamma,3}^{\prime 1-p_{3}}) + E_{\gamma,3}^{\prime 1-p_{3}}]^{1/(1-p_{3})}, & K_{2} < \xi_{1} < K_{3}\\[3pt] \relax
[\xi_{2}(E_{\gamma,5}^{\prime 1-p_{4}} - E_{\gamma,4}^{\prime 1-p_{4}}) + E_{\gamma,4}^{\prime 1-p_{4}}]^{1/(1-p_{4})}, & K_{3} < \xi_{1} < 1  \relax \nonumber
\end{array}
\right. 
\end{equation}\\

\section{Selection of electron for electron-photon scattering}
\label{AppendixB}
In this Appendix, we describe the algorithm to select an electron for scattering with a photon using the scattering probability $P_{scatt}$. We denote the angle between the propagation directions of a particular electron among $N_{e}$ electrons and the photon (already selected from the priority queue, see Section \ref{Sec2.5}) in the jet-comoving frame before scattering by $\theta_{e}^{\prime}$. The differential number of scatterings experienced by the photon in time $dt^{\prime}$ in jet-comoving frame is then given by,
\begin{equation}
dN_{scatt}^{\prime} = dn_{e}^{\prime} \sigma_{T}c (1 - \beta_{e}^{\prime}cos\theta_{e}^{\prime})dt^{\prime} \nonumber
\end{equation}
where, $dn_{e}^{\prime} = f(\beta_{e}^{\prime}, \Omega_{e}^{\prime}) d^{3}\beta_{e}^{\prime} d\Omega_{e}^{\prime}$ is differential element corresponding to the electron number density in the jet-comoving frame. $f(\beta_{e}^{\prime}, \Omega_{e}^{\prime})$ corresponds to the energy distribution of the electrons which is MB and $d^{3}\beta_{e}^{\prime} d\Omega_{e}^{\prime} = \beta_{e}^{\prime 2}d\beta_{e}^{\prime}\rm{sin}\:\theta_{e}^{\prime}d\theta_{e}^{\prime}d\phi_{e}^{\prime}$ is the differential element in the velocity space of the electrons. The probability of scattering between an electron and the photon is,
\begin{equation}
P_{scatt}(\beta_{e}^{\prime},\theta_{e}^{\prime}) \propto \frac{d\nu_{scatt}^{\prime}}{\beta_{e}^{\prime 2}d\beta_{e}^{\prime} d\Omega_{e}^{\prime}} = f(\beta_{e}^{\prime}, \theta_{e}^{\prime})\sigma_{T} c (1 - \beta_{e}^{\prime} \rm{cos}\:\theta_{e}^{\prime})  \nonumber
\end{equation}
where $d\nu_{scatt}^{\prime}$ is the differential frequency of electron-photon scattering.
$P_{scatt}$ is independent of $\phi_{e}^{\prime}$ because of azimuthal symmetry of the scattering event in the jet-comoving frame. Assuming that the electron distribution stays isotropic between scattering events
\begin{equation}
P_{scatt}(\beta_{e}^{\prime},\theta_{e}^{\prime}) \propto e^{-c \beta_{e}^{\prime 2}} (1 - \beta_{e}^{\prime} \rm{cos}\:\theta_{e}^{\prime})  \nonumber
\end{equation}
where, $c$ is a constant determined by the temperature of the electrons. The normalized probability can then be written as,
\begin{equation}
P_{scatt}(\beta_{e}^{\prime},\theta_{e}^{\prime}) = \frac{1}{4\pi \beta_{e}^{\prime 2}} (1 - \beta_{e}^{\prime}\rm{cos}\:\theta_{e}^{\prime}) \nonumber
\end{equation}
The cumulative distribution function corresponding to the above probability distribution is,
\begin{equation}
F_{scatt}(\beta_{e}^{\prime},\theta_{e}^{\prime}) = \frac{1}{2}\beta_{e}^{\prime}\left[(1- \rm{cos}\theta_{e}^{\prime}) + \frac{1}{4}\beta_{e}^{\prime}(\rm{cos}^{2}\theta_{e}^{\prime} - 1)\right] \nonumber
\end{equation}
which is zero for $\theta_{e}^{\prime} = 0$ and $\beta_{e}^{\prime}$ for $\theta_{e}^{\prime} = \pi$.
Next, we draw a random number $\xi_{3}$ between 0 and $N_{e} - 1$ and evaluate $\abs{\xi_{3} - N_{e}F_{scatt}(\theta_{e}^{\prime})}$ for all $N_{e}$ electrons. The electron selected for scattering with the photon is the one with minimum value of $\abs{\xi_{3} - N_{e}F_{scatt}(\theta_{e}^{\prime})}$.\\

\section{Pair production and annihilation}
\label{AppendixC}
The fraction of photons with sufficient energy, $E_{\gamma} \sim m_{e}c^{2}\Gamma \sim 1.5 \times 10^{5}$ keV, needed to create pairs in the jet is $f_{\nu} \sim 10^{-4}$ for $\tau_{in} = 4$ (see Fig. \ref{fig2}). Let $\eta$ be the fraction of photons that are close to the peak photon energy, $E_{\gamma,peak}=\Gamma h\nu_{sa}^{\prime}$, and within an energy range: $E_{\gamma,1}=\Gamma h(0.75\nu_{sa}^{\prime})$ to $E_{\gamma,2}=\Gamma h(1.25\nu_{sa}^{\prime})$. Then the number of photons with sufficient energy to produce pairs is, $N_{\gamma,MeV} \sim 10^{-4} \eta N_{\gamma,tot}$, where $N_{\gamma,tot} = 10^{8}$ is the total number of photons in the jet. 

The optical depth for pair production is, $\tau_{\gamma\gamma, MeV} \sim (N_{\gamma,MeV}\sigma_{\gamma\gamma,avg})/(4\pi R^2)$, where $\sigma_{\gamma\gamma,avg}=\int_{y_{min}}^{y_{max}}\sigma_{\gamma\gamma}(y)(f_{y}/y)dy\big{/}$ $\int_{y_{min}}^{y_{max}}(f_{y}/y)dy$ is the average pair production cross section with $y^2=\frac{1}{2}\frac{h\nu_{1}^{\prime}}{m_{e}c^2}\frac{h\nu_{2}^{\prime}}{m_{e}c^2}(1-\rm{cos\ }\theta)$.  Here, $\nu_{i}^{\prime}$ denotes the energy of the incoming photons and $\theta$ is the angle between them. Assuming isotropic photon distribution i.e. $\langle\rm{cos\ }\theta=0\rangle$, we have $y_{min}=1$, $y_{max} \sim 0.7(E_{\gamma,max}/\Gamma m_{e}c^2) \sim 50$ and $f_{y} \propto y^{-1.25}$ as the maximum possible photon energy $E_{\gamma,max} \sim 10^{7}$ keV from our simulation results. The average pair production cross section is then (\citealt{Poz83})

\begin{equation}
\sigma_{\gamma\gamma,avg} = \frac{\int_{1}^{50}\left(\frac{3}{8}\frac{\sigma_{T}}{y^{2}}\left[\left(2+\frac{2}{y^2}-\frac{1}{y^4}\right){\rm ln}(y+\sqrt{y^2-1})-\left(1+\frac{1}{y^2}\right)\left(1-\frac{1}{y^2}\right)^{1/2}\right]\right)y^{-2.25}dy}{\int_{1}^{50}y^{-2.25}dy} \sim 0.16\sigma_{T}
\end{equation}
Substituting $N_{\gamma,MeV}$ and $\sigma_{\gamma\gamma,avg}$, 
\begin{equation}
\tau_{\gamma\gamma,MeV} \sim 10^{-4}\eta \left(\frac{N_{\gamma,tot}}{N_{e,tot}}\right) \left(\frac{\sigma_{\gamma\gamma,avg} N_{e,tot}}{4\pi R^2}\right) \sim 10 \eta \times 0.16\tau_{e} \sim 1.6 \eta \tau_{e}
\label{taugamma}
\end{equation}
where we used $N_{\gamma,tot}/N_{e,tot} = 10^5$ and $\tau_{e} \sim (N_{e,tot}\sigma_{T})/(4\pi R^2)$. Therefore, $\tau_{\gamma\gamma,MeV} \lesssim 1$ is satisfied as long as $\eta \lesssim 1/6$ (for $\tau_{in} \sim 4$).

For the synchrotron seed spectrum of fast cooled electrons that we consider

\begin{equation}
\eta  \sim \frac{\int_{0.75\nu_{sa}^{\prime}}^{1.25\nu_{sa}^{\prime}} (f_{\nu}/\nu) d\nu}{\int_{\nu_{ac}^{\prime}}^{\nu_{m}^{\prime}} (f_{\nu}/\nu) d\nu} \sim \frac{\int_{0.75\nu_{sa}^{\prime}}^{\nu_{sa}^{\prime}}(\nu/\nu_{sa})^{3/8} d\nu + \int_{\nu_{sa}^{\prime}}^{1.25\nu_{sa}^{\prime}}(\nu/\nu_{sa})^{-3/2} d\nu}{\int_{0.01\nu_{sa}^{\prime}}^{\nu_{sa}^{\prime}}(\nu/\nu_{sa})^{3/8} d\nu + \int_{\nu_{sa}^{\prime}}^{500\nu_{sa}^{\prime}}(\nu/\nu_{sa})^{-3/2} d\nu} \sim 1/6
\end{equation}
which gives $\tau_{\gamma\gamma,MeV} \sim 1$ (from Equation \ref{taugamma}). It should be noted that here we make a conservative (although arbitrary) choice for the energy bin width $\Delta E_{\gamma}=\Gamma h(0.50\nu_{sa}^{\prime})$ as $\eta$ strictly corresponds to photons with energies very close to $\Gamma h\nu_{sa}^{\prime}$.

The optical depth for pair annihilation is, $\tau_{e^{-}e^{+}} \sim (N_{e^{-}e^{+}}\sigma_{e^{-}e^{+}})/(4\pi R^2) \sim (3/8)(N_{e^{-}e^{+}}/N_{e,tot})(\tau_{e}/\beta_{e}^{\prime})$, where $N_{e^{-}e^{+}}$ is the total number of pairs in the jet and $\sigma_{e^{-}e^{+}} \sim (3/8)(\sigma_{T}/\beta_{e}^{\prime})$ is the pair annihilation cross section. Equating the pair production and annihilation rates at equilibrium 
\begin{equation}
\frac{\sigma_{e^{-}e^{+}}N_{e^{-}e^{+}}\beta_{e}^{\prime}c}{4\pi R^2} = \frac{\sigma_{\gamma \gamma,avg}N_{\gamma,MeV}c}{4\pi R^2}
\end{equation}
which gives
\begin{equation}
N_{e^{-}e^{+}} \sim 4.2 N_{e,tot}\eta \sim 0.7 N_{e,tot}
\label{num_pairs}
\end{equation}
for $\eta \sim 1/6.$
Hence, the number of pairs $N_{e^{-}e^{+}}$ in the jet is always less than the total number of electrons $N_{e,tot}$. 

We have not explored $\tau_{in} \lesssim 4$ while evaluating $N_{e^{-}e^{+}}$ as the photons at such low optical depths do not experience enough scatterings for Comptonization to modify the seed photon spectrum appreciably. Although $f_{\nu}$ can be larger by a factor of $\gtrsim 5$ for $1 \lesssim \tau_{in} \lesssim 2$, the number density of pairs and thus the pair annihilation optical depth, $\tau_{e^{-}e^{+}} \sim \sigma_{e^{-}e^{+}}n_{e^{-}e^{+}}$, is also larger by the same factor. This increases the probability of the additional pairs getting annihilated very quickly and the number of pairs is comparable to that obtained in Equation \ref{num_pairs}. For larger values of $\tau_{in} \gtrsim 4$, $f_{\nu} \lesssim 10^{-5}$ which means that the number of pairs in the jet is even smaller. So, the effect of pairs can be ignored for the present work.

\end{appendix}

\label{lastpage}

\end{document}